\documentclass[%
  amsmath,amssymb,
]{revtex4-2}

\usepackage{graphicx}
\usepackage{bm}

\def\##1{\underline #1}
\def\=#1{\underline{\underline #1}}

\def\E{{\bf E}}
\def\e{{\bf e}}

\def\h{{\bf h}}

\def\barn{{\bar n}}

\def\epst{{\boldsymbol \epsilon}}

\def\Rot{{\cal R}}

\def\k{{\bf k}}
\def\r{{\bf r}}

\def\epst{{\boldsymbol \epsilon}}
\def\epstild{{\widetilde{\varepsilon}}}

\def\chit{{\boldsymbol\chi}}
\def\sigmab{{\boldsymbol\sigma}}
\def\epshat{\hat{\boldsymbol \epsilon}}
\def\epst{{\boldsymbol \epsilon}}
\def\barn{{\bar n}}
\def\Rot{{\cal R}}
\def\a{\alpha}

\usepackage{hyperref}
\usepackage{cleveref}
\hypersetup{
    colorlinks=true,
    linkcolor=blue,
    citecolor=blue,
    urlcolor=blue,
    breaklinks=true,
}

\begin{document}
\preprint{APS/123-QED}

\title{Reverse Circular Bragg Phenomenon}

\author{Martin W.\ McCall}
\email{m.mccall@imperial.ac.uk}
\affiliation{Blackett Laboratory, Department of Physics, Imperial College of Science, Technology and Medicine, Prince Consort Road, London SW7 2AZ, United Kingdom}
\author{Stefanos Fr.\ Koufidis}
\email{steven.koufidis20@imperial.ac.uk}
\affiliation{Blackett Laboratory, Department of Physics, Imperial College of Science, Technology and Medicine, Prince Consort Road, London SW7 2AZ, United Kingdom}

\date{\today}

\begin{abstract}
The axial propagation of circularly polarized light in an optically active structurally chiral medium is exactly solved via full electromagnetic analysis. Some symmetries of the system's characteristic matrix reveal new insights, which are confirmed by coupled wave theory. For extreme values of chirality, now accessible via metamaterials, a reverse circular Bragg resonance arises in the negative refraction regime where handedness reversal of counterpart modes occurs. A condition is identified under which optical activity offsets structural chirality, rendering the medium simply birefringent.
\end{abstract}

\maketitle

\section{Introduction}
Chirality, or the geometric property that an object cannot be super-imposed with its mirror image, is pervasive in nature \cite{Michelson1911}, and was recently studied in artificial meta-media for its potential applications in optoelectronics \cite{Kang2015} and sensing \cite{Yoo2019}. The  electromagnetic response of chiral media has two distinct origins. The first is via magneto-electric coupling, which in natural media arises from chirality at the molecular scale. The characteristic response of this form of chirality is optical activity, whereby the plane of linearly polarized light is rotated on transmission, or, equivalently, as circular birefringence, whereby right and left circular polarizations (RCP and LCP, respectively) propagate at different speeds. The other origin of electromagnetic chirality is via helical stacking of birefringent layers in structurally chiral media, such as cholesteric liquid crystals \cite{Lakhwani2012} or sculptured thin films \cite{Robbie1996}. Their signature response is the circular Bragg phenomenon, wherein co-handed
polarized light effectively experiences a periodic modulation of the refractive index by alternately sampling the two refractive indices associated with the two orthogonal axes progressively rotating along the direction of propagation. Then, monochromatic light is
strongly reflected when its wavelength in the medium matches the helical pitch. On the other hand, contra-handed polarized light, which helicity is opposite to the medium’s spatial handedness, propagates unaffected
seeing an average refractive index.

\par How to optically characterise media with either forms of chirality was discussed in \cite{Arteaga2016}, and the differentiation between structural and molecular chirality was explained in the conrm of polymer thin films in \cite{Campbell2020}. Either or both forms of chirality represent opportunities for manipulation of circularly polarized light that is attracting increasing attention in modern optoelectronics \cite{Han2018}. Such polarization selectivity can, e.g., enhance the throughput efficiency of OLED displays \cite{Li2015}, or enhance spin polarized electron transport in chiral organic semiconductors \cite{Yang2013}.

\par The asymmetrical response of structurally chiral media was studied in \cite{McCall2000, McCall2001, McCall2004, LakhtakiaBook2005, McCall2009}, where the problem of axial propagation was exactly solved via full electromagnetic analysis and approximately via coupled wave theory. The location of the Bragg resonances is sensitive to various parameters, and plays an important role in many applications in sensing \cite{Mackay2010sensing}. Indeed, in \cite{Lakhtakia2000}, the increment of the volumetric fraction shifted the Bragg resonance towards the blue, and an experiment conducted in \cite{Sorge2006} found a similar blue-shifting with increment of the deposition angle \cite{LakhtakiaBook2005}. Two opposite-sign lateral shifts at the edges of the Bragg zone of a nanocomposite structurally chiral medium were seen in \cite{Shirin2020}, and in \cite{Pursel2006}, a shift towards shorter wavelengths was associated to material changes occurring during annealing. The porosity of structurally chiral media offers a platform for fluid infiltration and a spectral shift, strongly dependent on the fluid’s refractive index, was demonstrated in \cite{Mackay2010}. This extended \cite{Lakhtakia2001}, where infiltration of void regions with a higher refractive index material augmented optical activity, which is approximately proportional to the square of the local linear birefringence \cite{Hodgkinson2000}.

\par  A natural question to address is the electromagnetic response of a medium that combines both structural and magneto-electric chirality. This appears to have been first examined numerically in \cite{Sherwin2003a, Sherwin2003b}, where the optical response of a chiral sculptured thin film, infiltrated by an isotropic chiral fluid, was studied. Apparently, the chiral fluid shifts the optical spectrum, linearly to its chirality, also affecting the peak reflectance and bandwidth. In this paper, the axial propagation of circularly polarized light in an optically active structurally chiral medium is solved analytically, and identification of some of the system's underlying symmetries reveals significant new insights. In particular, we demonstrate that additional resonances compared to those demonstrated in \cite{Sherwin2003a, Sherwin2003b} occur, as in \cite{Hodgkinson2004ambichiral, vanPopta2005}, but in a different kind of material-light interaction. When the optical activity parameter is extreme, eliciting negative refraction due to chirality, theoretically predicted in \cite{Pendry2004, Monzon2005} and experimentally realised in \cite{Zhang2009, Kafesak2009, Zheludev2009}, the  handedness and phase velocity of counterpart modes interchange. This reverse circular Bragg phenomenon is related to metamaterials with giant optical activity \cite{Zarifi2012}, and is fundamentally distinct to the handedness reversal phenomenon due to the permittivity and permeability being simultaneously negative of \cite{Lakhtakia2003}. 

\par The paper is organized as follows: in Sec.\ \ref{Constitutive Relations}, constitutive relations for optically active structurally chiral media are formed and in Sec.\ \ref{Full electromagnetic analysis for axial propagation}, the problem is exactly solved via full electromagnetic analysis. In Sec.\ \ref{Optical response}, the optical spectrum of a slab of the considered medium is investigated and the novel resonances are demonstrated. Finally, a condition under which optical activity precisely offsets structural chirality, rendering the medium simply birefringent, is identified for the first time in Sec.\ \ref{Optical activity offsetting structural chirality}.

\section{Constitutive Relations}
\label{Constitutive Relations}
\subsection{Constitutive relations for optically active media}
According to the Drude-Born-Fedorov model \cite{BohrenBook2003}, the temporal frequency domain constitutive relations for optical activity in a bi-isotropic reciprocal medium are
\begin{align}
    \mathbf{D}&=\varepsilon_0\left(\varepsilon\mathbf{E}+i\alpha \eta_0\mathbf{H}\right)\,, \label{DBF Standard Relations 1}
    \\
    \mathbf{B}&=\mu_0\left[-i\left({\alpha}/{\eta_0}\right)\mathbf{E}+\mu\mathbf{H}\right]\,, \label{DBF Standard Relations 2}
\end{align}
where $\mathbf{E}$, $\mathbf{B}$ are the fundamental electromagnetic fields and $\mathbf{D}$, $\mathbf{H}$ are the excitation fields. The free-space permittivity, permeability, and impedance are $\varepsilon_0$, $\mu_{0}$, and $\eta_0=\left({{\mu_0}/{\varepsilon_0}}\right)^{1/2}$, respectively, the relative permittivity is $\varepsilon$, and the relative permeability is $\mu$. The chirality parameter $\alpha$ measures the distance (in wavelengths) after which the $\mathbf{E}$-vector of a linearly polarized wave completes a full rotation. Defining the auxiliary fields  $\mathbf{h}=\eta_0\mathbf{H}$, $\mathbf{b}=\left({\eta_0}/{\mu_0}\right)\mathbf{B}$, and $\mathbf{d}=\varepsilon_0^{-1}\mathbf{D}$, with the same dimensions as $\mathbf{E}$, Eqs.\ \eqref{DBF Standard Relations 1} and \eqref{DBF Standard Relations 2} become, respectively,
\begin{align*}
    \mathbf{d}&= \varepsilon\mathbf{E}+i\alpha\mathbf{h}\,, 
    \\
    \mathbf{b}&=-i\alpha\mathbf{E}+\mu\mathbf{h}\,. 
\end{align*}
Maxwell’s macroscopic source-free curl relations, under the $\exp{\left(-i\omega t\right)}$ harmonic convention, are written as
\begin{align}
     \nabla\times\mathbf{E}&=k_0\alpha\mathbf{E}+i k_0\mu\mathbf{h}\,, \label{Maxwell 1}
     \\
     \nabla\times\mathbf{h}&=-ik_0\varepsilon\mathbf{E}+k_0\alpha\mathbf{h}\,, \label{Maxwell 2}
\end{align}
where $k_0= (\varepsilon_0\mu_0)^{1/2}\omega$ is the free-space wave number. For plane wave propagation along $z$, with a unit vector $\hat{\bf z}$, combining Eqs.\ \eqref{Maxwell 1} and \eqref{Maxwell 2} yields
\begin{equation}
    \frac{{ \rm d}^2\mathbf{E}}{{ \rm d} z^2}+2\alpha k_0\frac{ \rm d}{{ \rm d}z}\left(\hat{\mathbf{z}}\times\mathbf{E}\right)+k_0^2\left(\mu\varepsilon-\alpha^2\right)\mathbf{E}=\mathbf{0}\,,
    \label{Helmholtz Wave Equation}
\end{equation}
which is the Helmholtz wave equation describing axial propagation in an isotropic optically active medium.

\subsection{Constitutive relations for structurally chiral media}
A structurally chiral medium is modelled by a dielectric tensor of the form 
\begin{equation*}
    \epst = \Rot\cdot \epshat \cdot \Rot^{-1}\,,
    \label{Tensor of Structural Chirality}
\end{equation*}
where $\epshat$ is the background (i.e., static) dielectric tensor and in Cartesian coordinates, 
\begin{equation*}
    \Rot=\left(\begin{matrix}\cos{\left(pz\right)}&-h\sin{\left(pz\right)}&0\\h\sin{\left(pz\right)}&\cos{\left(pz\right)}&0\\0&0&1\\\end{matrix}\right)
    \label{eq Rmat}
\end{equation*}
describes the periodic rotation of the eigenaxes of $\epst{}$ along and about the $z$-axis with a spatial period $L_{ {p}}=2\pi/p$ \cite{McCall2009}. For $h=+1$, the eigenaxes of $\epst$ rotate in a right-handed (RH) sense while for $h=-1$, the rotation is regarded as left-handed (LH). The transverse projection of $\mathcal{R}$ can be expressed as \cite{McCall2000}
\begin{equation}
          \mathcal{R}_\bot=\frac{1}{2}\sigmab e^{i pz}+\frac{1}{2}\sigmab^\ast e^{-i pz}\,,
\label{Transverse R}
\end{equation} 
where
\begin{equation*}
   \sigmab=\left(\begin{matrix}1&hi\\-hi&1\\\end{matrix}\right)\,.
\end{equation*}

\par Denoting as $\epst_{{\rm ref}}$ the reference tensor, with Cartesian components $\epst_{{\rm ref}}={\rm  diag}\left(\varepsilon_{ {a}},\varepsilon_{ {b}},\varepsilon_{ {c}}\right)$, the background tensor accounts for the orientation of the principal axes as
\begin{equation*}
    \epshat =\chit\cdot\epst_{{\rm ref}}\cdot{\chit}^{-1}\,, 
\label{eq epsilon par}
\end{equation*}
where
\begin{equation*}
    \chit=\left(\begin{matrix}\cos{\chi}&0&-\sin{\chi}\\0&1&0\\\sin{\chi}&0&\cos{\chi}\\\end{matrix}\right)\,,
\end{equation*}
and the so-called rise angle $\chi$ tilts the eigenaxes in the $x$-$z$ plane. It is straightforward to show that the  projection of $\epshat$ in the $x$-$y$ plane takes the diagonal form \cite{McCall2009}
\begin{equation}
    \epshat_\perp=\left(\begin{matrix}\epstild&0\\0&\varepsilon_{ {b}}\\\end{matrix}\right)\,, 
\label{eq epsilon matrix}
\end{equation}
where 
\begin{equation*}
    \widetilde{\varepsilon}=\frac{\varepsilon_{ {a}}\varepsilon_{ {c}}}{\varepsilon_{ {a}}\sin^2\chi+\varepsilon_{ {c}}\cos^2\chi}\,.
\end{equation*}
It is convenient to re-express the tensor of Eq.\ \eqref{eq epsilon matrix} as
\begin{equation*}
    \epshat_\perp =\bar{\varepsilon}\  {\mathbb{I}}+\delta\varepsilon\ {\mathbb{J}}\,,
\label{eq decompose}
\end{equation*}
where
\begin{align}
    \bar{\varepsilon}&={\left(\widetilde{\varepsilon}+\varepsilon_{ {b}}\right)}/{2}\,, \label{Average Epsilon}
    \\
    \delta\varepsilon&={\left(\widetilde{\varepsilon}-\varepsilon_{ {b}}\right)}/{2}\,, \nonumber
\end{align}
${\mathbb{I}}$ is the $2\times2$ identity, and  ${\mathbb{J}}={\rm diag}\left(1,-1\right)$. Overall,
\begin{equation}
    \epst_\perp \left(z,h\right) = \mathcal{R}_\bot \cdot \epshat_\bot \cdot \mathcal{R}_\bot^{-1}\,,
\label{eq epsilon transverse}
\end{equation}
is the $z$-dependent tensor describing structural chirality, which also depends on the handedness parameter $h$, characterising the medium, via Eq.\ \eqref{Transverse R}.

\subsection{Constitutive relations for optically active structurally chiral media}
As we are interested in axial propagation in an anisotropic medium that combines optical activity with structural chirality, the scalar $\varepsilon$ appearing in Eq.\ \eqref{Helmholtz Wave Equation}, for the transverse component of the electric field $\mathbf{E}_{\bot}$, is replaced by the tensor $\epst_\perp$ of Eq.\ \eqref{eq epsilon transverse}, presumed to be inhomogeneous along the $z$-direction only. Hence, such a Helmholtz wave equation, shall be describing axial propagation in optically active structurally chiral media.

\section{Full electromagnetic analysis for axial propagation}
\label{Full electromagnetic analysis for axial propagation}
Considering the transverse components of the fields, Eqs.\ \eqref{Maxwell 1} and \eqref{Maxwell 2} can be written in a matrix notation as
\begin{equation}
    \frac{ \rm d}{\rm d z}\left(\times\right)\left(\begin{matrix}\mathbf{E}_\bot\\\mathbf{h}_\bot\\\end{matrix}\right)=i k_0\left(\begin{matrix}-i {\a}\mathbb{I}&\mu\mathbb{I}\\-\epst_\perp &-i {\a}\mathbb{I}\\\end{matrix}\right)\left(\begin{matrix}\mathbf{E}_\bot\\\mathbf{h}_\bot\\\end{matrix}\right)\,,
\label{eq Max Sys}
\end{equation}
where $\left(\times\right)$ has Cartesian components $ \left(\begin{matrix}0&-1\\1&0\\\end{matrix}\right)$ and acts on both $\mathbf{E}_\bot$ and $\mathbf{h}_\bot$. The system of Eq.\ \eqref{eq Max Sys} can be rendered autonomous via the Oseen transformation \cite{Oseen1993}
\begin{subequations}
\begin{align}
     \mathbf{e}&=\mathcal{R}_\bot^{-1}\cdot\mathbf{E}_\bot\,,  \label{Oseen Transformation 1}
     \\
     \widetilde{\mathbf{h}}&=\mathcal{R}_\bot^{-1}\cdot\mathbf{h}_\bot\,, \label{Oseen Transformation 2}
\end{align}
\end{subequations}
under the convention of \cite{ McCallBook2015}. Via Eqs.\ \eqref{Oseen Transformation 1} and \eqref{Oseen Transformation 2}, the two equations of the system in Eq.\ \eqref{eq Max Sys} become
\begin{align*}
     \frac{ \rm d}{{ \rm d}z}\left(\times\right)\mathbf{e}&=-\mathcal{R}_{\bot}^{-1}\frac{ {\rm d}\mathcal{R}_{\bot}}{{ {\rm d}}z}\left(\times\right)\mathbf{e}+ak_0\mathbf{e}+ik_0\mu\widetilde{\mathbf{h}}\,,
     \\
     \frac{ \rm d}{{ \rm d}z}\left(\times\right)\widetilde{\mathbf{h}}&=-\mathcal{R}^{-1}\frac{ {\rm d} \mathcal{R}_{\bot}}{{ \rm d}z}\left(\times\right)\widetilde{\mathbf{h}}-ik_0\epshat_\bot\cdot\mathbf{e}+ak_0\ \widetilde{\mathbf{h}}\,, 
\end{align*}
and it is easy to show that they can be simplified to
\begin{equation}
    \frac{ \rm d}{\rm d z}\left(\begin{matrix}\mathbf{e}\\\widetilde{\mathbf{h}}\\\end{matrix}\right)=\left(\times\right)\left(\begin{matrix}-hp-k_0{\a}&-i k_0\mu\\i k_0\epshat_\bot&-hp-k_0{\a}\\\end{matrix}\right)\left(\begin{matrix}\mathbf{e}\\\widetilde{\mathbf{h}}\\\end{matrix}\right)\,.
\label{eq Oseen System}
\end{equation}
In terms of the $\mathbf{G}=\left(e_{x},\ e_{y},\ \widetilde{h}_{x},\ \widetilde{h}_{y}\right)^T$ components of the Oseen transformation fields, where $T$ denotes transpose, Eq.\ \eqref{eq Oseen System} is equivalently written as
\begin{equation}
    \frac{{ \rm d} \mathbf{G}}{{ \rm d}z}=\mathbf{F}\cdot\mathbf{G}\,,
    \label{General Equation}
\end{equation}
where the components of the characteristic matrix are
\begin{equation}
    \mathbf{F}=\left(\begin{matrix}
0 & hp+k_0{\a} & 0 & i k_0\mu \\
-hp-k_0{\a} & 0 & -i k_0\mu & 0 \\
0 & -i k_0 \epsilon_{  b} & 0 & hp+k_0{\a} \\
i k_0 \tilde{\epsilon} & 0 & -hp-k_0{\a} & 0 
\end{matrix} \right)\,.
\label{Characteristic Matrix F}
\end{equation}

\par
\begin{figure}[ht]
\centering
\includegraphics[width=0.5\linewidth]{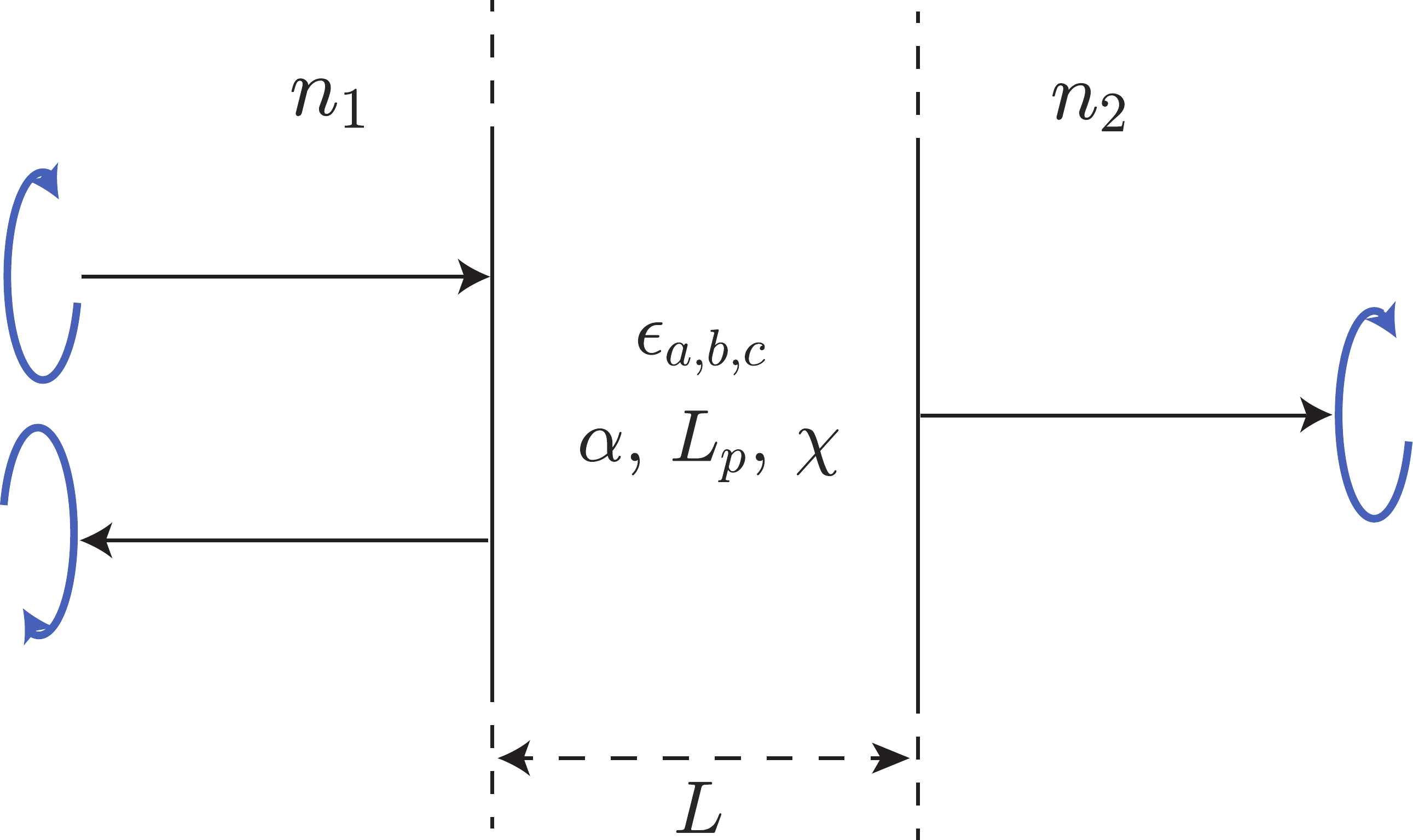}
\caption{A length $L=6$ \textmu m slab of a right-handed ($h=+1$) optically active structurally chiral medium, stacked between two non index-matched isotropic dielectrics with refractive indices $n_1=1$ and $n_2=2$. The arrows indicate the incident, reflected, and transmitted circular polarizations. The medium's parameters are: $\epsilon_{  a}=3.2+0.02i$, $\epsilon_{  b}=2.9+0.02i$, $\epsilon_{  c}=2.8+0.02i$, $\mu=1$, $L_{ {p}}=300$ nm, and $\chi=30^\circ$.}
\label{Figure 1}
\end{figure}

 The exact solution to the autonomous system of Eq.\ \eqref{General Equation} can be used to construct the axially propagating fields via Eqs.\ \eqref{Oseen Transformation 1} and \eqref{Oseen Transformation 2}, from which the reflection and transmission coefficients of polarized light incident to a slab of the considered medium may be computed. Indeed, assuming a slab of length $L$, as illustrated in Fig.\ \ref{Figure 1}, the solution to Eq.\ \eqref{General Equation} for the transverse components of the fields is \cite{McCall2009}
\begin{equation*}
    \left(\begin{matrix}{\bf E}_\bot\\{{\bf h}}_\bot\\\end{matrix}\right)_{z=L}=\left(\begin{matrix}\mathcal{R}_{\bot}&\mathbf{0}\\\mathbf{0}&\mathcal{R}_{\bot}\\\end{matrix}\right)e^{\mathbf{F}L}\left(\begin{matrix}{\bf E}_\bot\\{{\bf h}}_\bot\\\end{matrix}\right)_{z=0}\,,
    \label{actual E and H}
\end{equation*}
where ${\bf 0}$ is the $2 \times 2$ null matrix, and $e^{\mathbf{F}L}$ is a well-defined matrix exponential. For $a_{  x,y}$, $r_{  x,y}$, and $t_{  x,y}$ the amplitudes of the incident, reflected, and transmitted electric fields in Cartesian coordinates, respectively, field-matching at both interfaces of Fig.\ \ref{Figure 1} requires
\begin{align*}
     \left(\begin{matrix}{\bf E}_\bot\\{\bf h}_\bot\\\end{matrix}\right)_{z=0}&=\left(\begin{matrix}a_{  x}+r_{  x}\\a_{  y}+r_{  y}\\-n_1\left(a_{  y}-r_{  y}\right)\\n_1\left(a_{  x}-r_{  x}\right)\\\end{matrix}\right)\,,
     \\
     \left(\begin{matrix}{\bf E}_\bot\\{\bf h}_\bot\\\end{matrix}\right)_{z=L}&=\left(\begin{matrix}t_{  x}\\t_{  y}\\-n_2t_{  y}\\n_2t_{  x}\\\end{matrix}\right)\,.
\end{align*}
The linear reflection and transmission coefficients are defined as
\begin{align*}
     \left(\begin{matrix}r_{  x}\\r_{  y}\\\end{matrix}\right)&=\left(\begin{matrix}r_{  xx}&r_{  xy}\\r_{  yx}&r_{  yy}\\\end{matrix}\right)\left(\begin{matrix}a_{  x}\\a_{  y}\\\end{matrix}\right)\,,
     \\
      \left(\begin{matrix}t_{  x}\\t_{  y}\\\end{matrix}\right)&=\left(\begin{matrix}t_{  xx}&t_{  xy}\\t_{  yx}&t_{  yy}\\\end{matrix}\right)\left(\begin{matrix}a_x\\a_{y}\\\end{matrix}\right)\,, 
\end{align*}
respectively, and converting to a circular basis yields
\begin{align*}
    \left(\begin{matrix}r_{  LL}&r_{  LR}\\r_{  RL}&r_{  RR}\\\end{matrix}\right)&=\frac{1}{2}\left(\begin{matrix}1&i\\1&-i\\\end{matrix}\right)\left(\begin{matrix}r_{  xx}&r_{  xy}\\r_{  yx}&r_{  yy}\\\end{matrix}\right)\left(\begin{matrix}1&1\\i&-i\\\end{matrix}\right)    \,,
    \\
    \left(\begin{matrix}t_{  LL}&t_{  LR}\\t_{  RL}&t_{  RR}\\\end{matrix}\right)&=\frac{1}{2}\left(\begin{matrix}1&-i\\1&i\\\end{matrix}\right)\left(\begin{matrix}t_{  xx}&t_{  xy}\\t_{  yx}&t_{  yy}\\\end{matrix}\right)\left(\begin{matrix}1&1\\i&-i\\\end{matrix}\right)\,.
\end{align*}
\par The intensity reflectances are then $R_{i,j}=|r_{i,j}|^2$, while the intensity transmittances $T_{i,j}=\left({n_2}/{n_1}\right)|t_{i,j}|^2$, with $\{i, j\}=\{  R, L\}$ indicating reflection/transmission, of the $i$ polarization for incident $j$ polarization. 

\section{Optical response}
\label{Optical response}

\subsection{Regular circular Bragg phenomenon}
\label{Regular circular Bragg phenomenon}
In the absence of optical activity, $\alpha = 0$ and the characteristic matrix $\mathbf{F}$ of Eq.\ \eqref{Characteristic Matrix F} describes axial propagation through a usual structurally chiral medium. It is well-established that the resulting optical spectrum shows the circular Bragg phenomenon if the condition $\lambda_0^{ {\rm Br}} = {\rm  Re}(\barn) L_{ {p}}$, where $\barn = (\bar{\varepsilon}\mu)^{1/2}$ with $\bar{\varepsilon}$ defined in Eq. \eqref{Average Epsilon}, is met (see, e.g., \cite{McCall2000, McCall2001, McCallBook2015, McCall2009}). In this case, circularly polarized light that is co-handed with the medium will be strongly reflected, whereas contra-handed light will be transmitted.

\par If the structurally chiral medium is now infiltrated by a chiral fluid, i.e., if $\alpha\neq0$, it will also become optically active. Comparing Eq.\ \eqref{General Equation} to  \cite[Eq.\ (42)]{McCall2009}, it is evident that combining optical activity with structural chirality is mathematically described by a \emph{linear perturbation} to the characteristic matrix of a simple uninfiltrated structurally chiral medium, which depends on both the chirality $\alpha$ and the handedness $h$, namely
\begin{equation}
    p \rightarrow p+\left({\alpha}/{h}\right)k_0\,.
    \label{General Linear Perturbation}
\end{equation}

\par In an optically active structurally chiral medium, the perturbation of Eq.\ \eqref{General Linear Perturbation} will result in a shift of the Bragg resonance, determined from
\begin{equation}
     \left.\lambda_0^{ {\rm Br}}\right|_{\rm  RH (LH)}^{\rm  RCP (LCP)} =\frac{{\rm  Re}(\barn)}{{L_{ {p}}}^{-1}+ {\alpha}/{ \left.\lambda_0^{ {  Br}}\right|_{\rm  RH (LH)}}}\,. 
     \label{wavelngth shift}
\end{equation}
Hence, in a right-handed optically active structurally chiral medium, RCP light will be resonant at
\begin{equation}
    \left.\lambda_0^{ {\rm Br}}\right|_{ \rm RH}^{\rm RCP} = L_{ {p}} \left[{\rm  Re}(\barn) - \alpha\right]\,,\ \ \alpha < {\rm  Re}(\barn)\,,
    \label{Regular CBP for RH}
\end{equation}
whereas in a left-handed optically active structurally chiral medium, LCP light will be resonant at \begin{equation}
    \left.\lambda_0^{ {\rm Br}}\right|_{ \rm LH}^{ \rm LCP} = L_{ {p}} \left[{\rm  Re}(\barn)  + \alpha\right]\,,\ \ \alpha > -{\rm  Re}(\barn)\,,
    \label{Regular CBP for LH}
\end{equation}
where we have also stated the conditions that $\alpha$ and $\barn$ must satisfy so that the wavelengths remain positive. These considerations quantitatively explain the spectral shifts observed numerically in \cite{Sherwin2003a, Sherwin2003b}, since our system in Eq.\ \eqref{General Equation} is a special case of \cite[Eq.\ (18)]{Sherwin2003a} describing infiltration by a general anisotropic chiral fluid. 

\par
\begin{figure}[ht]
\centering
\includegraphics[width=0.5\linewidth]{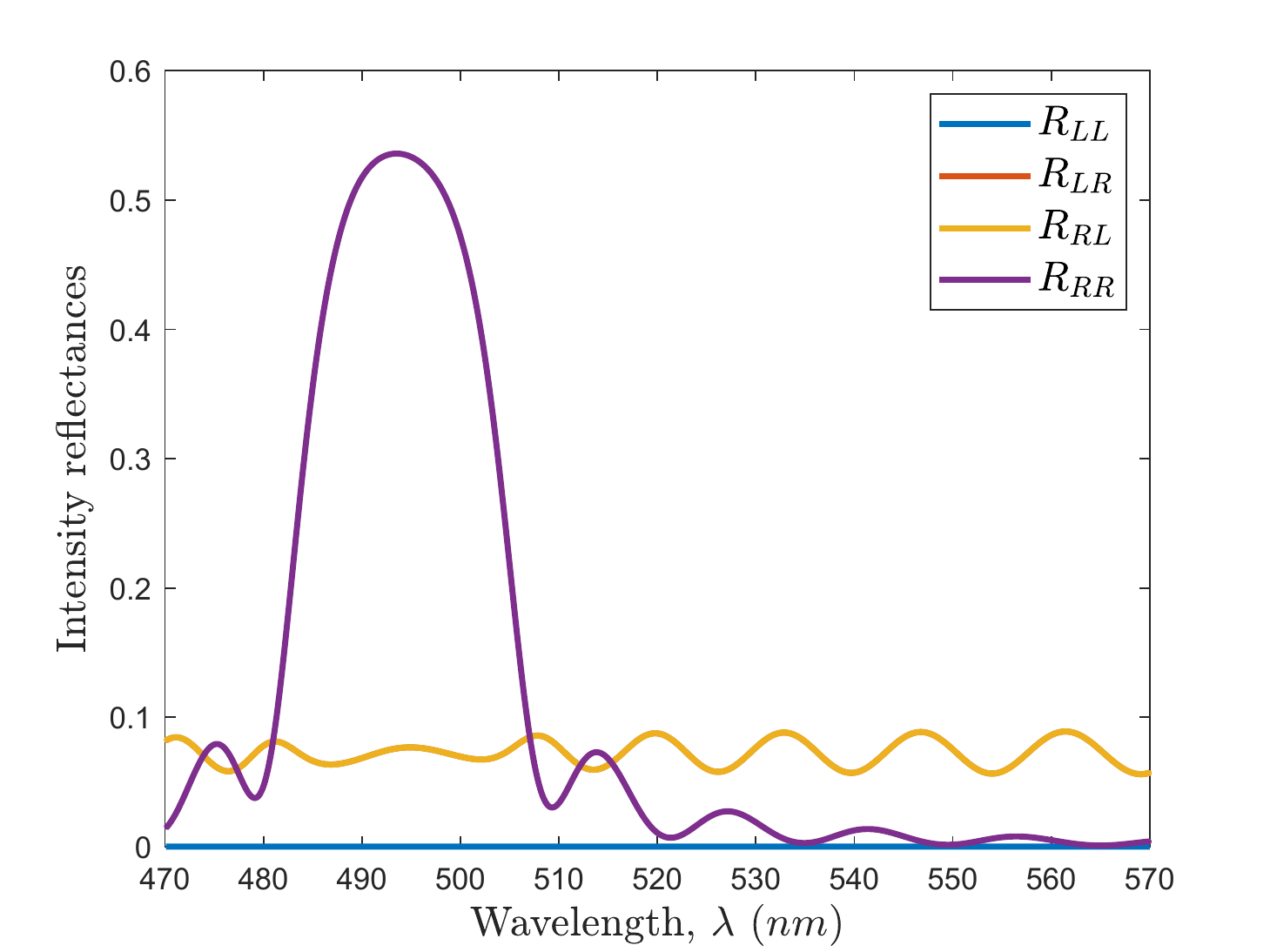}
\caption{Intensity reflectances for a slab of an optically active structurally chiral medium for a physical value of chirality, ${\alpha}=0.084$. Optical activity shifts the Bragg regime, linearly to the chirality. The parameters are those of Fig.\ \ref{Figure 1}.}
\label{Figure 2}
\end{figure}

\begin{figure}[ht]
\centering
\includegraphics[width=0.5\linewidth]{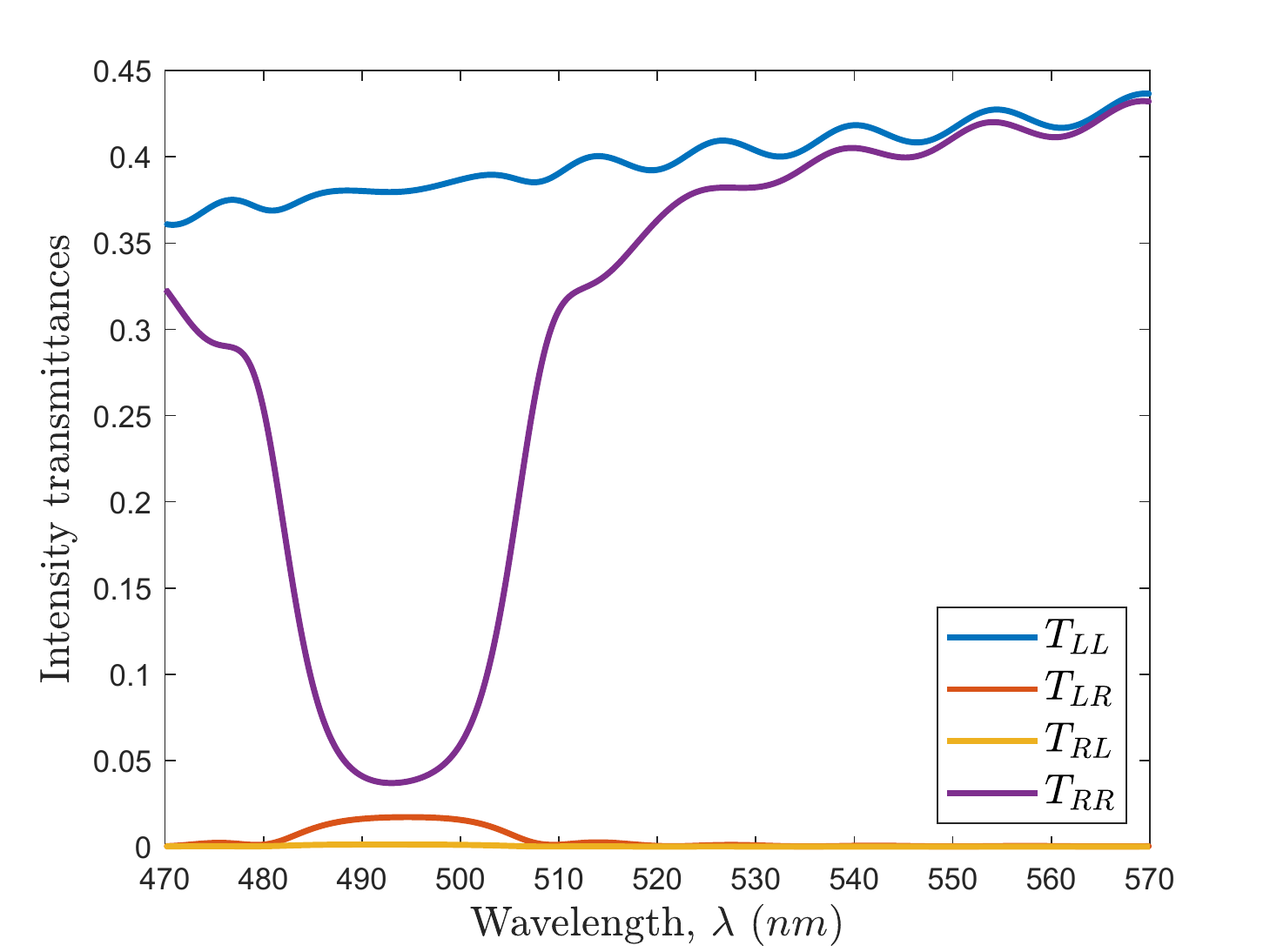}
\caption{Intensity transmittances for a slab of an optically active structurally chiral medium for a physical value of chirality, ${\alpha}=0.084$. In the presence of absorption, the transmittance responds asymmetrically to the reflectance and $R+T\neq1$. The parameters are those of Fig.\ \ref{Figure 1}.}
\label{Figure 3}
\end{figure}

 To demonstrate these resonances, we consider a sculptured thin film, which porous nature makes it ideal to be infiltrated by an optically active chiral fluid, e.g., a D-Glucose solution ${  C}_\mathbf{6}{  H}_{\mathbf{12}}{  O}_\mathbf{6}\cdot{  H}_\mathbf{2}{  O}$. In \cite{Mohammadi2016}, the real part of the specific chirality of such a solution is $\left[\alpha\right]\approx1.37\cdot{10}^{-6}\,{  cm^3/g}$, while the imaginary part is almost zero (at $21.5^{\circ} \ {  C}$). For a concentration close to saturation, $C=1.8 \  {{g}/{cm^3}}$ and the solution's chirality is $\alpha=2.46\cdot{10}^{-6}$. For the purpose of demonstration, we set ${\alpha}=0.084$ which is a reasonably high but not extreme value (almost 6 times less than the chirality used in \cite{Sherwin2003a}), lying within $-\bar{n}<{\a}<\bar{n}$. The intensity reflectances and transmittances, for the setup in Fig.\ \ref{Figure 1}, can be seen in Figs.\ \ref{Figure 2} and \ref{Figure 3}, respectively. 

\par Evidently, optically active structurally chiral media exhibit a circular Bragg phenomenon for co-handed polarised light. In the absence of optical activity the central Bragg wavelength would be ${\lambda}_0^{\rm Br}=519.1$ nm while in its presence, it is shifted towards the blue at $\left(\lambda_0^{ {\rm Br}}\right)^{\prime}=493.9$ nm  for a right-handed medium, as per Eq. \eqref{Regular CBP for RH}, or towards the red at $\left(\lambda_0^{ {\rm Br}}\right)^{\prime\prime}=544.3$ nm for a left-handed one, as per Eq.\ \eqref{Regular CBP for LH}, by $|\delta\lambda|=|L_{  p}\alpha|=25.2$ nm. This spectral shift characterizes both the medium and the chiral fluid, offering a platform for sensing either the concentration or the specific chirality of a solution (see, e.g., \cite{Mackay2010sensing}).

\subsection{Reverse circular Bragg phenomenon}
The characteristic matrix of Eq.\ \eqref{Characteristic Matrix F} is generally a function of the medium's parameters, i.e., $\mathbf{F}\left(\varepsilon_{  b},\tilde{\varepsilon},\mu;k_0;p,h,\alpha\right)$. Should we perform the transformation
\begin{equation}
    \left(\varepsilon_{  b},\tilde{\varepsilon},\mu;k_0;p,h,\alpha\right)\rightarrow\left(\varepsilon_{  b},\tilde{\varepsilon},\mu;k_0;-p,h,-\alpha\right)\,,
    \label{FEMA matrix transformation 1}
\end{equation}
so that the perturbation of Eq.\ \eqref{General Linear Perturbation} becomes $p\rightarrow-p-\left({\alpha}/{h}\right)k_0$, we effectively reverse the handedness via $p$ and not via $h$, which is fixed for a particular medium. For
$\varepsilon_{  b}$, $\tilde{\varepsilon}$, and $\mu$ unchanged, it turns out that an underlying symmetry of the characteristic matrix $\mathbf{F}$ is
\begin{equation}
    \mathbf{F}\left(p,h,\alpha\right)=\mathcal{H}\cdot\mathbf{F}\left(-p,h,-\alpha\right)\cdot\mathcal{H}\,, 
    \label{FEMA matrix transformed}
\end{equation}
where $\mathcal{H}={\rm  diag}\left(1,-1,-1,1\right)$, with $\mathcal{H}^{-1}=\mathcal{H}$. We now return to Eq.\ \eqref{General Equation} which via Eq.\ \eqref{FEMA matrix transformed} is written as
\begin{equation*}
    \frac{{ \rm d} \left[\mathbf{G}\left(p,h,\alpha\right)\right]}{{ \rm d} z}=\left[\mathcal{H}\cdot\mathbf{F}\left(-p,h,-\alpha\right)\cdot\mathcal{H}\right]\cdot\mathbf{G}\left(p,h,\alpha\right)\,,
\end{equation*}
and implies that
\begin{equation}
    \mathbf{G}\left(p,h,\alpha\right)=\mathcal{H}\cdot\mathbf{G}\left(-p,h,-\alpha\right)\,.
    \label{FEMA Symmetry}
\end{equation}
Then, according to \cite[Eq.\ (18)]{Lakhtakia2003}, Eq.\ \eqref{FEMA Symmetry} entails that the indices of the reflection and transmission coefficients are interchanged ($  R \leftrightarrow L$), i.e., nominally RCP light will be now LCP and vice versa. Hence, the transformation of Eq.\ \eqref{FEMA matrix transformation 1}, creates the illusion that the medium's handedness is reversed, although $h$ is fixed.

\par
\begin{figure}[ht]
\centering
\includegraphics[width=0.5\linewidth]{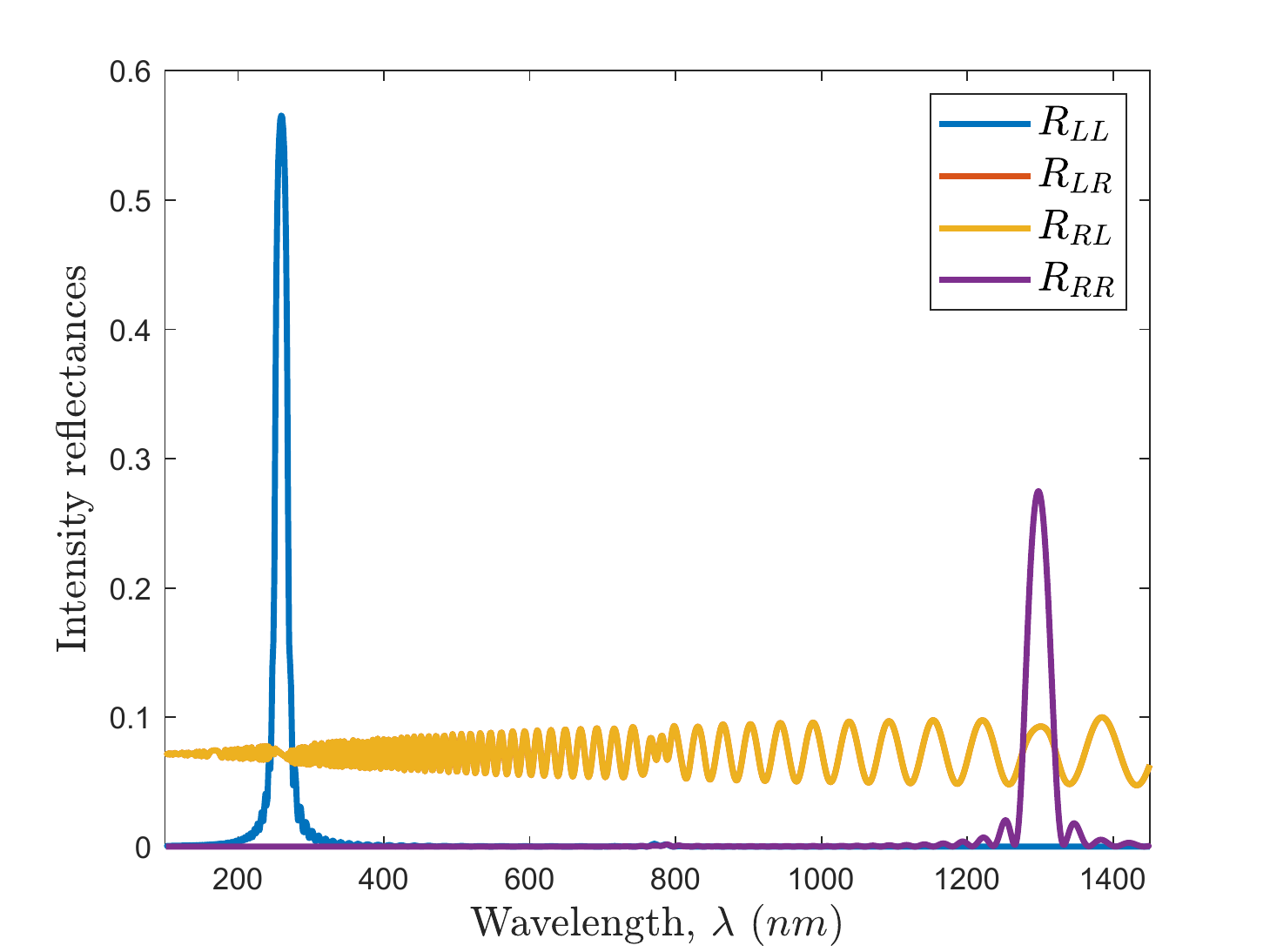}
\caption{Intensity reflectances for a slab of an optically active structurally chiral medium for an extreme value of chirality, ${\alpha}=-1.5{\rm  Re}\left(\bar{n}\right)$. Two resonances arise: one, regular, for RCP light at $\left.\lambda_0^{ {\rm Br}}\right|_{ \rm RH}^{\rm RCP} =1.29$ \textmu m and one, reverse, for LCP light at $\left.\lambda_0^{ {\rm Br}}\right|_{ \rm RH}^{\rm LCP} =259.55$ nm. The parameters are those of Fig.\ \ref{Figure 1}.}
\label{Figure 4}
\end{figure}

\begin{figure}[ht]
\centering
\includegraphics[width=0.5\linewidth]{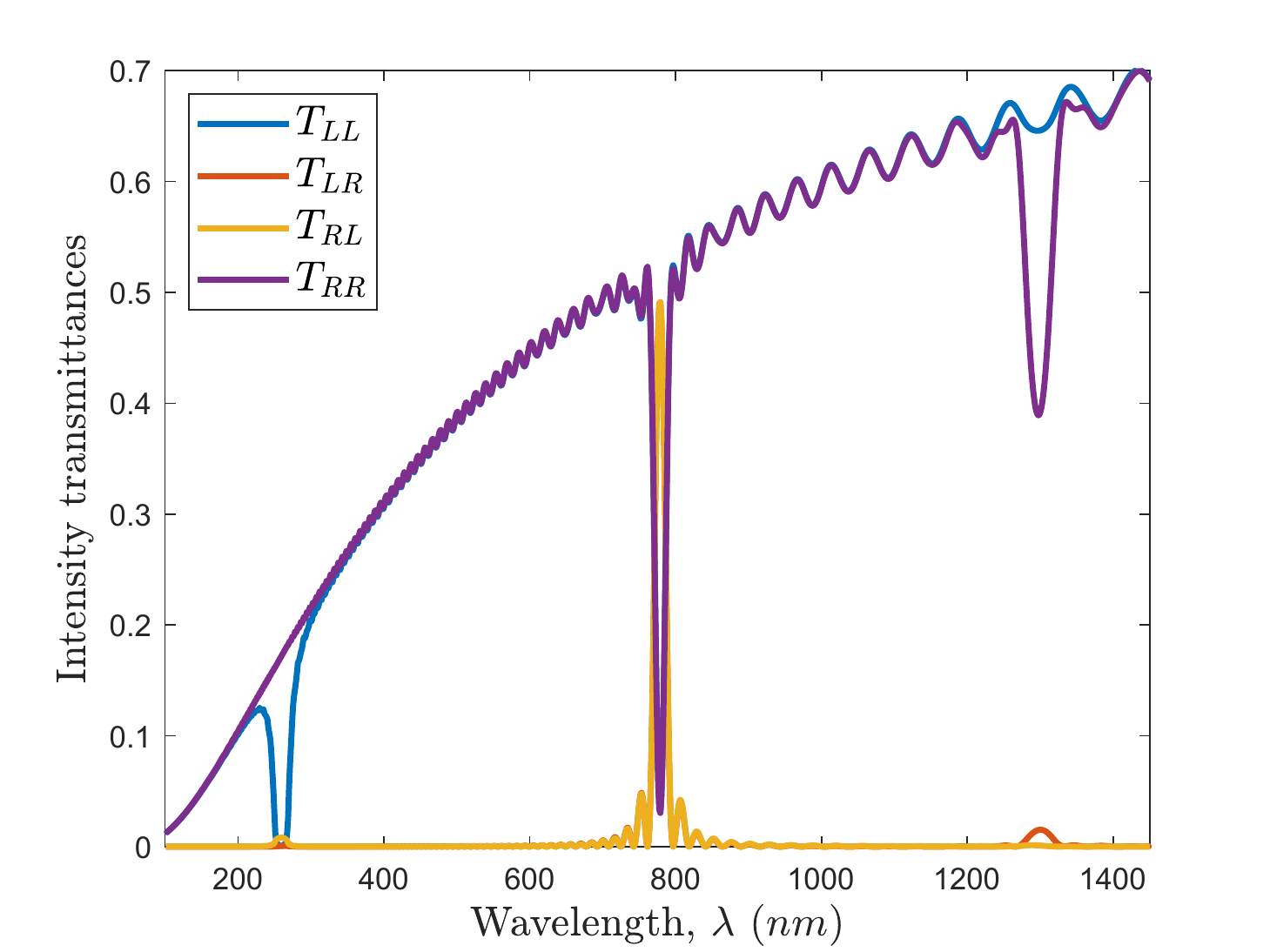}
\caption{Intensity transmittances for a slab of an optically active structurally chiral medium for an extreme value of chirality, ${\alpha}=-1.5{\rm  Re}\left(\bar{n}\right)$. The ramp behaviour is associated to increased absorption for increased optical thickness. At a wavelength exactly between the two resonances, $\lambda_0^{  c}=778.6$ nm, optical activity offsets structural chirality and the medium is simply birefringent. The parameters are those of Fig.\ \ref{Figure 1}.}
\label{Figure 5}
\end{figure}

 We now apply the transformation of Eq.\ \eqref{FEMA matrix transformation 1} to Eq.\ \eqref{wavelngth shift}, which explicitly links the resonance of RCP (LCP) light to a right-handed (left-handed) medium. Following the arguments above, the resulting resonance in a right-handed medium will be for LCP light at
\begin{equation}
    \left.\lambda_0^{ {\rm Br}}\right|_{ \rm RH}^{\rm LCP} = -L_{ {p}}[{\rm  Re}(\barn)+\alpha]\,, \ \ \alpha < -{\rm  Re}(\barn)\,,
    \label{Reverse CBP for RH}
\end{equation}
while in a left-handed medium, for RCP light, at 
\begin{equation}
    \left.\lambda_0^{ {\rm Br}}\right|_{ \rm LH}^{\rm RCP} = -L_{ {p}}[{\rm  Re}(\barn)-\alpha]\,, \ \ \alpha >{\rm  Re}(\barn)\,.
    \label{Reverse CBP for LH}
\end{equation}
Remarkably, in a right-handed medium, the condition for the resonance occurrence in Eq.\ \eqref{Regular CBP for RH}, $\alpha<{\rm  Re}(\barn)$, is \emph{still satisfied} when the condition of Eq.\ \eqref{Reverse CBP for RH} is met. Consequently, for $\alpha <- {\rm  Re}(\barn)$, two resonances are expected, one corresponding to light that is co-handed with the medium being reflected, and another, new resonance, that back-scatters light that is nominally \emph{contra}-handed with the medium. Similarly, for a left-handed medium the condition of Eq.\ \eqref{Reverse CBP for LH}, $\alpha >{\rm  Re}(\barn)$, simultaneously satisfies the condition of Eq.\ \eqref{Regular CBP for LH}. This back-scatters RCP light in a left-handed medium. These resonances are signatures of a ``\emph{Reverse Circular Bragg Phenomenon}''. The optical response of the slab of Fig.\ \ref{Figure 1} for an extreme value of chirality is demonstrated in Figs.\ \ref{Figure 4} and \ref{Figure 5}.

\par The requirement $|\alpha|>{\rm  Re}(\barn)$ signifies extreme values of optical activity, equivalent to a rotation of the electric field vector within a wavelength in the medium. This has been identified in \cite{Pendry2004, Monzon2005} as the negative refraction due to chirality regime and in \cite{Mackay2005}, modes with negative phase velocity in isotropic chiral media were demonstrated. This is directly related to the handedness reversal phenomenon of \cite{Lakhtakia2003}, where both the permittivity and permeability of a structurally chiral medium become negative. Accessing such large values of chirality has been achieved in metamaterials at THz frequencies in \cite{Zhang2009} and at GHz frequencies in \cite{Kafesak2009, Zheludev2009}, without, however, requiring the permittivity and permeability to be \emph{simultaneously negative}. As we demonstrate in Appendix \ref{Review of Optical Activity}, these resonances exploit the exchange of handedness occurring for extreme optical activity, where the handedness of the co-handed forward propagating mode swaps with the handedness of the contra-handed backward propagating one. The existence of these remarkable resonances is further confirmed via coupled wave theory in Appendix \ref{Coupled wave theory description of optical activity with structural chirality}. 

\section{Optical activity offsetting structural chirality}
\label{Optical activity offsetting structural chirality}
A further interesting observation can be made on Eq.\ \eqref{Characteristic Matrix F}. In fact, when $k_0\alpha=-hp$ (n.b. $\alpha$ can take either sign), or in terms of wavelength at
\begin{eqnarray}
    \lambda_0^{  c}&\underset{h=+1}{=}&  \frac{1}{2}\left(\left.\lambda_0^{ {\rm Br}}\right|^{\rm LCP}_{ \rm RH} + \left.\lambda_0^{ {\rm Br}}\right|^{\rm RCP}_{ \rm RH}\right)= -L_{ {p}}\alpha\,,\nonumber\\
    \nonumber\\
    &\underset{h=-1}{=}&\frac{1}{2} \left(\left.\lambda_0^{ {\rm Br}}\right|^{\rm LCP}_{ \rm LH} + \left.\lambda_0^{ {\rm Br}}\right|^{\rm RCP}_{ \rm LH}\right)= L_{ {p}}\alpha\,,
     \label{eq intermediate wavelength}
\end{eqnarray}
which lies exactly in between the two resonances of Fig.\ \ref{Figure 5} and is equal to the chirality induced spectral shift,  optical activity \emph{precisely offsets} structural chirality. At this wavelength, the fields $\mathbf{e}$, $\widetilde{\mathbf{h}}$, which are twisting along with the helix, experience a linearly birefringent medium. The forward propagating modes are then $\mathbf{e}_{1}=\left(\exp{\left(i{\tilde n}k_0z\right)},\ 0\right)^T$ and $\mathbf{e}_{2}=\left(0, \ \exp{\left(i{n_b}k_0z\right)}\right)^T$, where $\tilde{n}=\left({\tilde\varepsilon}\mu \right)^{{1}/{2}}$ and $n_{  b}=\left(\varepsilon_{ {b}}\mu \right)^{{1}/{2}}$. When transformed back using Eqs.\ \eqref{Oseen Transformation 1} and \eqref{Oseen Transformation 2}, we get
\begin{subequations}
\begin{align}
    {\bf E}_1 &= e^{i{\tilde n}k_0z}\left(\begin{matrix}\cos{\left(pz\right)}\\h\sin{\left(pz\right)}\end{matrix}\right)\,, \label{eq corkscrew mode 1}
     \\
    {\bf E}_2 &= e^{i{n_{  b}}k_0z}\left(\begin{matrix}-h\sin{\left(pz\right)}\\\cos{\left(pz\right)}\end{matrix}\right)\,,
    \label{eq corkscrew mode 2}
\end{align}
\end{subequations}
which correspond to linearly polarized fields, in Cartesian coordinates, that rotate with the eigenaxes of the structurally chiral medium. The total transverse field is
\begin{equation}
     {\bf E}_\perp = A_1 {\bf E}_1 + A_2 {\bf E}_2\,,
    \label{Total field eq corkscrew modes}
\end{equation}
where $A_{1,2}$ are the amplitudes of the electric fields. Interestingly, ${\bf E}_1$ and ${\bf E}_2$ are at once \emph{orthogonal but co-handed}, taking the same chirality as the structurally chiral medium. Converting the expression in Eq.\ \eqref{Total field eq corkscrew modes} to an equivalent in a circular basis, we obtain
\begin{equation}
    \left(\begin{matrix}E_{  L}^+\\E_{  R}^+\\\end{matrix}\right)_{z=L}=\left(\begin{matrix}e^{i k_0\left(\bar{n}+\alpha\right) z}\cos{\left(k_0\tilde{\tilde{n}} z\right)}&i e^{i k_0\left(\bar{n}+\alpha\right)z}\sin{\left(k_0\tilde{\tilde{n}} z\right)}\\i e^{-i k_0\left(\bar{n}-\alpha\right) z}\sin{\left(k_0\tilde{\tilde{n}} z\right)}&e^{-i k_0 \left(\bar{n}-\alpha\right) z}\cos{\left(k_0\tilde{\tilde{n}} z\right)}\\\end{matrix}\right)\left(\begin{matrix}E_{  L}^+\\E_{  R}^+\\\end{matrix}\right)_{z=0}\,,
\label{eq Ofs FEMA}
\end{equation}
with $\tilde{\tilde{n}}={\left(\tilde{n}-n_{ {b}}\right)}/{2}$. This represents a continuous exchange between circular states as light propagates through a birefringent medium.

\par
\begin{figure}[ht]
\centering
\includegraphics[width=0.5\linewidth]{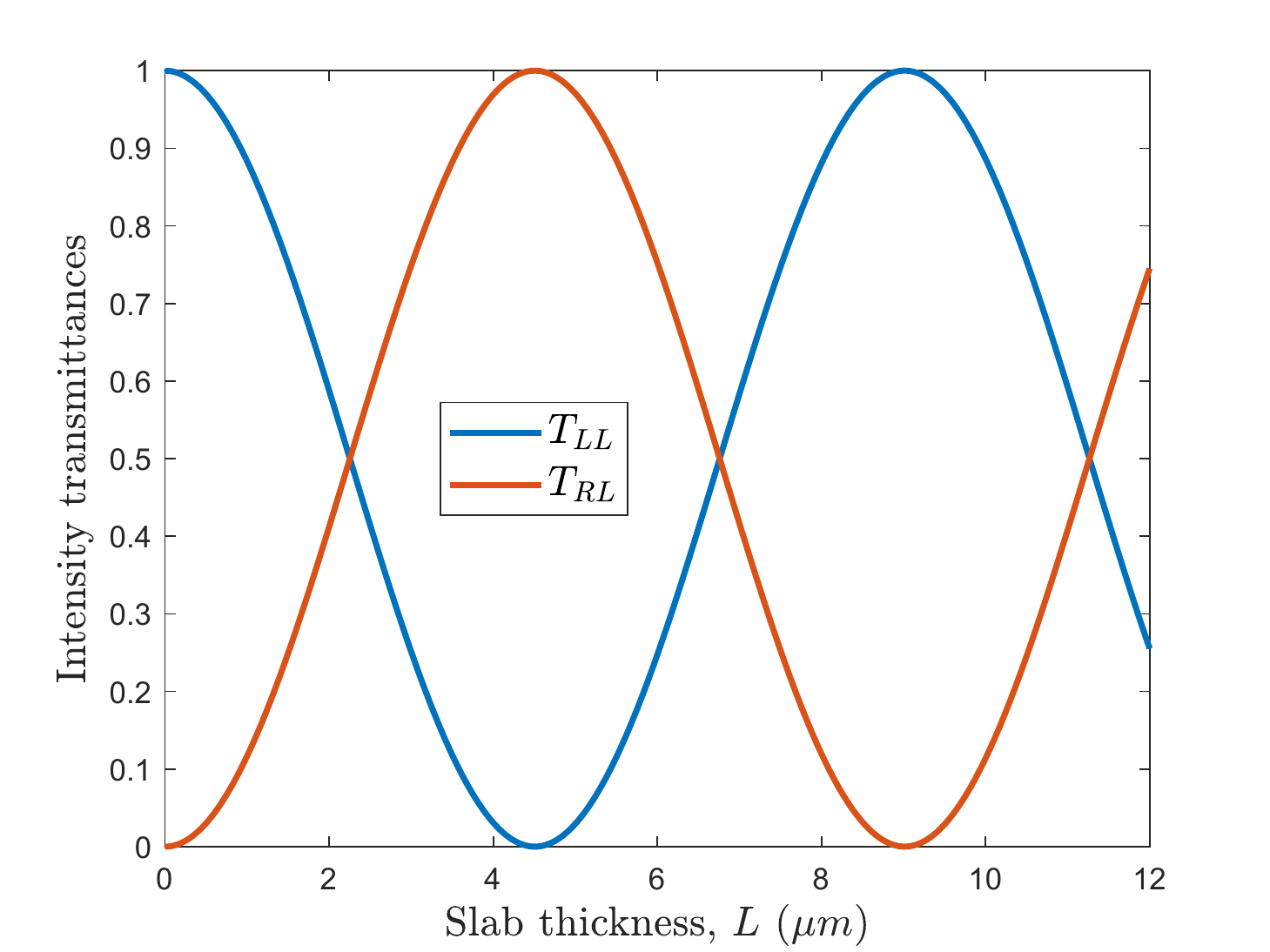}
\caption{Periodic exchange of circular states as a function of the slab's thickness at $\lambda_0^{  c}$ (cf. Fig.\ \ref{Figure 5}) as per Eq.\ \eqref{eq Ofs FEMA}. The parameters are those of Fig.\ \ref{Figure 1}.}
\label{Figure 6}
\end{figure}

 At $\lambda_0^{  c}$, plotting the transmittances of Fig.\ \ref{Figure 1}, accessed via Eq.\ \eqref{eq Ofs FEMA}, as functions of the slab's thickness in Fig.\ \ref{Figure 6}, we see that the energy between $T_{  LL}$ ($T_{  RR}$) and $T_{  RL}$ ($T_{  LR}$) is exchanged. In this regime, optical activity counteracts structural chirality, effectively ``unwrapping" the medium which light now sees as a simply birefringent, with refractive indices $\tilde{n}$ and $n_{b}$. Moving towards shorter wavelengths, the medium is ``wrapped” again but this time with an opposite handedness so that a reverse circular Bragg phenomenon is possible. 

\section{Conclusions}
The axial propagation of circularly polarized light in optically active structurally chiral media was exactly solved via full electromagnetic analysis, and some underlying symmetries of the system's characteristic matrix led to additional resonances, previously unknown. For chirality comparable to the average refractive index, in the negative refraction regime, the handedness of the co-handed forward propagating mode swaps with the handedness of the contra-handed backward propagating one, bringing the latter into resonance. At another wavelength, lying exactly between the two resonances, optical activity exactly offsets structural chirality, rendering the medium simply birefringent. There, circular polarizations degenerate to linear orthogonal, and co-handed, rotating at the same rate as the structurally chiral medium's eigenaxes. These results might prove of particular significance in a range of optoelectronics and sensing applications.

\section*{Appendices}
\appendix 

\section{Review of optical activity}
\label{Review of Optical Activity}
The eigenvalues of the matrix $\mathbf{S}$ on the right Eq.\ \eqref{eq Max Sys}, for $\epst_\perp$ reverted to $\varepsilon$, are $\lambda_\pm=
-i\left(\alpha\pm\barn\right)$, each with multiplicity two, with corresponding eigenvectors ${\bf u}_{\pm}=\left({\bf v},\mp{i {\bf v}}/{\eta}\right)^T$, where $\eta=\left({\mu}/{\varepsilon}\right)^{{1}/{2}}$ and ${\bf v}\in\mathbb{C}^2$. Then,
\begin{equation*}
    \mathbf{D}=\left(\begin{matrix}\lambda_+\mathbb{I}&\mathbf{0}\\\mathbf{0}&\lambda_-\mathbb{I}\\\end{matrix}\right)=\mathbf{U}^{-1}\cdot\mathbf{S}\cdot\mathbf{U}\,,
\end{equation*}
where
\begin{equation*}
    \mathbf{U}=\left(\begin{matrix}\mathbb{I}&\mathbb{I}\\-{i}/{\eta}\mathbb{I}&{i}/{\eta\mathbb{I}}\\\end{matrix}\right)\,.
\end{equation*}
For fields propagating along the $z$-axis, with an $\exp{\left(ikz\right)}$ dependence, Eq.\ \eqref{eq Max Sys} is expressed as
\begin{equation}
    k(\times)\left(\begin{matrix}{\bf Q}_1\\{\bf Q}_2\\\end{matrix}\right)=\left(\begin{matrix}-i\gamma_+\mathbb{I}&\mathbf{0}\\\mathbf{0}&-i\gamma_-\mathbb{I}\\\end{matrix}\right)\left(\begin{matrix}{\bf Q}_1\\{\bf Q}_2\\\end{matrix}\right)\,,
    \label{diagonal form}
\end{equation}
where $\gamma_\pm=\ k_0\left(\alpha\pm\barn\right)$, and the Beltrami fields are
\begin{equation*}
    \left(\begin{matrix}{\bf Q}_1\\{\bf Q}_2\\\end{matrix}\right)=\mathbf{U}^{-1}\left(\begin{matrix}\bf{E}_\bot\\\bf{h}_\bot\\\end{matrix}\right)=\frac{1}{2}\left(\begin{matrix}{\bf E}_\bot+i\eta\bf{h}_\bot\\{\bf E}_\bot-i\eta\bf{h}_\bot\\\end{matrix}\right)\,.
    \label{definition of Q_plus and Q_minus}
\end{equation*}
Then, equation \eqref{diagonal form} yields
\begin{equation*}
    \left(\begin{matrix}i\gamma_+&-k\\k&i\gamma_+\\\end{matrix}\right)\left(\begin{matrix}Q_{1x}\\Q_{1y}\\\end{matrix}\right)=0\,,
\end{equation*}
with eigenvalues $k_{1\pm}=\pm\gamma_+$, and
\begin{equation*}
   \left(\begin{matrix}i\gamma_-&-k\\k&i\gamma_-\\\end{matrix}\right)\left(\begin{matrix}Q_{2x}\\Q_{2y}\\\end{matrix}\right)=0\,, 
\end{equation*}
with eigenvalues $k_{2\pm}=\pm\gamma_-$. 

\par For $|\alpha| < {\rm  Re}\left(\bar{n}\right)$, the modes propagating along $+z$ have wave numbers $k_{1+}$ and $k_{2-}$. The corresponding eigenvectors propagate according to 
\begin{align*}
     {{\bf Q}_{1+}}&={Q_{1+}}\frac{e^{i k_0(\barn+\alpha)z}}{\sqrt2}\left(\begin{matrix}1\\i\\\end{matrix}\right)\,, 
     \\
     {{\bf Q}_{2+}}&={Q_{2+}}\frac{e^{i k_0(\barn-\alpha)z}}{\sqrt2}\left(\begin{matrix}1\\-i\\\end{matrix}\right)\,,
\end{align*}
where $Q_{1+}$ and $Q_{2+}$ are constants. If ${Q}_{2+}={0}$, then ${\bf E}_\bot={\bf E}_{1+}={\bf Q}_{1+}$, and we have
\begin{equation*}
    {\rm  Re}
   \left({\bf E}_{1+}\right)=
   \frac{|Q_{1+}|e^{-k_0{\rm  Im}(\barn)z}}{\sqrt2}
  \left(\begin{matrix}\cos{\left[\gamma_{+}z+{  arg}\left(Q_{1+}\right)\right]}\\-\sin{\left[\gamma_{+}z+{  arg}\left(Q_{1+}\right)\right]}\\\end{matrix}\right)\,,
    \label{LH Electric Field plus}
\end{equation*}
corresponding to an electric field that describes a left-handed helix in space. Similarly, if ${Q}_{1+}={0}$, then ${{\bf E}_\bot={\bf E}_{2+}={\bf Q}}_{2+}$ and
\begin{equation*}
    {\rm  Re}\left({\bf E}_{2+}\right)
    =\frac{|Q_{2+}|e^{-k_0{\rm  Im}(\barn)z}}{\sqrt2}
   \left(\begin{matrix}\cos{\left[-\gamma_{-}z+{  arg}\left(Q_{2+}\right)\right]}\\\sin{\left[-\gamma_{-}z+{  arg}\left(Q_{2+}\right)\right]}\\\end{matrix}\right)\,,
\end{equation*}
corresponding to a field describing a right-handed helix in space. The modes propagating along $-z$ have wave numbers $k_{1-}$ and $k_{2+}$ with corresponding eigenvectors
\begin{align*}
    {{\bf Q}_{1-}}&={Q_{1-}}\frac{e^{-i k_0(\barn+\alpha)z}}{\sqrt2}\left(\begin{matrix}1\\i\\\end{matrix}\right)\,,
    \\
    {{\bf Q}_{2-}}&={Q_{2-}}\frac{e^{-i k_0(\barn-\alpha)z}}{\sqrt2}\left(\begin{matrix}1\\-i\\\end{matrix}\right)\,.
\end{align*}
The fields, found by setting $Q_{2-}=0$ and $Q_{1-}=0$ are, respectively,
\begin{align*}
    {\rm  Re}
   \left({\bf E}_{1-}\right)&= \frac{|Q_{1-}|e^{k_0{\rm  Im}(\barn)z}}{\sqrt2} \left(\begin{matrix}\cos{\left[\gamma_{+}z+{  arg}\left(Q_{1-}\right)\right]}\\\sin{\left[\gamma_{+}z+{  arg}\left(Q_{1-}\right)\right]}\\\end{matrix}\right)\,,
   \\
   {\rm  Re}\left({\bf E}_{2-}\right)
    &=\frac{|Q_{2-}|e^{k_0{\rm  Im}(\barn)z}}{\sqrt2}
     \left(\begin{matrix}\cos{\left[-\gamma_{-}z+{  arg}\left(Q_{2-}\right)\right]}\\-\sin{\left[-\gamma_{-}z+{  arg}\left(Q_{2-}\right)\right]}\\\end{matrix}\right)\,.
\end{align*}

\par When $\alpha > {\rm  Re}\left(\barn\right)$ ($\alpha <- {\rm  Re}\left(\barn\right)$), the direction of phase advance for ${\bf E}_{2+}$ (${\bf E}_{1+}$) changes from being positive to negative, so that ${\bf E}_{2+}$ (${\bf E}_{1+}$) describes a wave whose phase propagates along $-z$.  Moreover ${\rm  Re}({\bf E}_{2+})$ (${\rm  Re}({\bf E}_{1+})$) changes from describing a right-handed (left-handed) helix to a left-handed (right-handed) helix. Similarly, the direction of phase advance for ${\bf E}_{2-}$ (${\bf E}_{1-}$) changes from being negative to positive, so that ${\bf E}_{2-}$ (${\bf E}_{1-}$) describes a wave whose phase propagates along $+z$ and ${\rm  Re}({\bf E}_{2-})$ (${\rm  Re} ({\bf E}_{1-})$) changes from describing a left-handed (right-handed) helix to a right-handed (left-handed) helix. While the direction of phase propagation changes, the Poynting vector does not \cite{McCall2009negativerefraction}.

\section{Coupled wave theory description of optical activity with structural chirality}
\label{Coupled wave theory description of optical activity with structural chirality}

Reverting the tensor $\epst$ to a scalar $\varepsilon$, Eq.\ \eqref{Helmholtz Wave Equation} will be describing optical activity without structural chirality. The transverse electric field of plane waves is given by
\begin{align}
    \mathbf{E}_\bot=&\left(A_{  L}^+e^{i k_0\left(\bar{n}+{\a}\right)z}+A_{  R}^-e^{-i k_0\left(\bar{n}-{\a}\right)z}\right){\bf Q}_{1}
    +\left(A_{  R}^+e^{i k_0\left(\bar{n}-{\a}\right)z}+A_{  L}^-e^{-i k_0\left(\bar{n}+{\a}\right)z}\right){\bf Q}_{2}\,,
\label{eq Ansatz}
\end{align}
in a circular basis, where
\begin{equation}
    {\bf Q}_{1}=\frac{1}{\sqrt2}\left(\begin{matrix}1\\i\\\end{matrix}\right)\ \ {\rm  and}\ \ {\bf Q}_{2}=\frac{1}{\sqrt2}\left(\begin{matrix}1\\-i\\\end{matrix}\right)\,. \label{eq Q1 and 2}
\end{equation}
For $|\alpha| < \barn$, the amplitudes of Eq.\ \eqref{eq Ansatz} are appropriately notated, whereas for $|\alpha| > \barn$, the phase velocity and handedness of some modes reverse - see Appendix \ref{Review of Optical Activity}.

\par The amplitudes are assumed to be slowly varying functions of $z$. Substituting Eq.\ \eqref{eq Ansatz}, along with Eqs.\ \eqref{Transverse R}, \eqref{eq epsilon matrix}, and \eqref{eq epsilon transverse} into Eq.\ \eqref{Helmholtz Wave Equation}, retaining potentially phase-matched terms, and resolving along $\mathbf{Q}_{1}$ and $\mathbf{Q}_{2}$ of Eq.\ \eqref{eq Q1 and 2} yields
\begin{align}
     \frac{{ \rm d} A_{  L}^+}{{ \rm d} z}e^{i k_0\left(\bar{n}+{\a}\right)z}-\frac{{ \rm d} A_{  R}^-}{{ \rm d} z}e^{-i k_0\left(\bar{n}-{\a}\right)z}=
     i \kappa A_{  R}^+e^{i \left[k_0\left(\bar{n}-{\a}\right)-2hp\right]z}+i \kappa A_{  L}^-e^{-i \left[k_0\left(\bar{n}+{\a}\right)+2hp\right]z}
    \label{eq Term Resolve Q1}
\end{align}
and
\begin{align}
     \frac{{ \rm d} A_{  R}^+}{{ \rm d} z}e^{i k_0\left(\bar{n}-{\a}\right)z}-\frac{{ \rm d} A_{  L}^-}{{ \rm d} z}e^{-i k_0\left(\bar{n}+{\a}\right)z}=
    i \kappa A_{  L}^+e^{i \left[k_0\left(\bar{n}+{\a}\right)+2hp\right]z}+i \kappa A_{  R}^-e^{-i \left[k_0\left(\bar{n}-{\a}\right)-2hp\right]z}\,.
    \label{eq Term Resolve Q2}
\end{align}
The coupling constant is $\kappa=\pi\delta\bar{n}/\lambda_0$, where $\delta\bar{n}=|\tilde{n}-n_{ {b}}|$ is the local linear birefringence. The absorption is almost constant over the range of $\delta \barn$, thus $\kappa \in \mathbb{R}$. 

\par For different phase-matching scenarios, Eqs.\ \eqref{eq Term Resolve Q1} and \eqref{eq Term Resolve Q2} can be distilled into
\begin{equation}
    \frac{{ \rm d}}{{ \rm d}z}\left(\begin{array}{c}A_{  L}^+\\A_{  L}^-\\A_{  R}^+\\A_{  R}^-\end{array}\right) = i \kappa\left(\begin{array}{cccc}0&e^{-i \delta_{  L}z}&e^{-i \delta_{  c} z}&0\\
                                         -e^{i \delta_{  L}z}&0&0&-e^{i \delta_{  c} z}\\
                                         e^{i \delta_{  c} z}&0&0&e^{-i \delta_{  R} z}\\
                                         0&-e^{-i \delta_{  c} z}&-e^{i \delta_{  R}z}&0\end{array}\right)
                                         \left(\begin{array}{c}A_{  L}^+\\A_{  L}^-\\A_{  R}^+\\A_{  R}^-\end{array}\right)\,,
    \label{CWT System}
\end{equation}
and the various detunings are given by
\begin{align*}
    \delta_{  R}&=2k_0\left(\bar{n}-\alpha\right)-2hp\,,  
    \\
    \delta_{  L}&=2k_0\left(\bar{n}+\alpha\right)+2hp\,, 
    \\
    \delta_{  c} &= 2k_0\alpha+2hp\,.
\end{align*}

\par For a right-handed medium, the on-resonance condition for RCP light, ${\rm  Re}(\delta_{  R})=0$, corroborates Eq.\  \eqref{Regular CBP for RH}. For a left-handed medium the ${\rm  Re}(\delta_{  L})=0$ condition for LCP light corroborates Eq.\ \eqref{Regular CBP for LH}. Finally, setting $\delta_{  c} = 0$, we obtain $k_0\alpha =-hp$, corroborating Eq.\ \eqref{eq intermediate wavelength} and corresponding to the modes given by Eqs.\ \eqref{eq corkscrew mode 1} and \eqref{eq corkscrew mode 2}. 

\par For each identified resonance, coupled wave theory provides a pair of readily solved equations from which the relevant $z$-dependent amplitudes are deduced. Then, the total transverse electric field may be reconstructed via Eq.\ \eqref{eq Ansatz} so that a medium's optical spectrum can be calculated as in \cite{McCall2009}. For $\delta_{  R,L} \approx 0$, the amplitudes evolutions are given by the solutions to Eqs. \eqref{CWT System}
\begin{equation*}
    \left(\begin{array}{c}A_{  R,L}^+\\A_{  R,L}^-\end{array}\right)_{z=L} =
    \left(\begin{array}{cc}p_{  R,L}^{+}&q_{  R,L}^{+}\\q_{  R,L}^{-}&p_{  R,L}^{-}\end{array}\right) \left(\begin{array}{c}A_{  R,L}^+\\A_{  R,L}^-\end{array}\right)_{z=0}\,,
    \label{CWE Solution RH Medium}
\end{equation*}
where 
\begin{align*}
    p_{  R,L}^{\pm}\left(z\right)=& e^{\mp \frac{i\delta_{  R,L}}{2}z}\left[\cosh{\left(\Delta_{  R,L} z\right)}\pm i \frac{\delta_{  R,L} }{2\Delta_{  R,L}}\sinh{\left(\Delta_{  R,L} z\right)}\right]\,,
   \\
    q_{  R,L}^{\pm}\left(z\right)=&\pm i e^{\mp \frac{i\delta_{  R,L}}{2}z} \frac{\kappa}{\Delta_{  R,L}} \sinh{\left(\Delta_{  R,L} z\right)}\,, 
\end{align*}
with $\Delta_{  R,L} = \left[\kappa^2-{\left({\delta_{  R,L}}/{2}\right)}^2\right]^{1/2}$. 

\par For $\delta_{  c}\approx 0$, the solution is slightly different. In fact,
\begin{equation*}
    \left(\begin{array}{c}A_{  L}^+\\A_{  R}^+\end{array}\right)_{z=L} =
    \left(\begin{array}{cc}\tilde{p}^{+}&\tilde{q}\\\tilde{q}&\tilde{p}^{-}\end{array}\right) \left(\begin{array}{c}A_{  L}^+\\A_{  R}^+\end{array}\right)_{z=0}\,,
    \label{copropagatingsolution}
\end{equation*}
where
\begin{align*}
    \tilde{p}^{\pm}\left(z\right) &= e^{\mp i\frac{\delta_{ {c}}}{2}z}\left[\cos{\left(\Delta_{ {c}} z\right)}\pm i \frac{\delta_{ {c}}}{2\Delta_{ {c}}}\sin{\left(\Delta_{ {c}} z\right)}\right]\,,
    \\
    \tilde{q}\left(z\right) &= i e^{\mp i\frac{\delta_{ {c}}}{2}z}\frac{\kappa }{\Delta_{ {c}}} \sin{\left(\Delta_{ {c}} z\right)}\,,
    \label{P and Q tilde definition}
\end{align*}
with $\Delta_{ {c}}= \left[\kappa^2+{\left({\delta_{  c}}/{2}\right)}^2\right]^{1/2}$. The circular components of the electric fields are then
\begin{equation}
    \left(\begin{array}{c}E_{  L}^+\\E_{  R}^+\end{array}\right)_{z=L} =e^{i k_0\barn z}
    \left(\begin{array}{cc}e^{-i hpz}\tilde{p}^{+}&e^{-i hpz}\tilde{q}\\e^{i hpz}\tilde{q}&e^{i hpz}\tilde{p}^{-}\end{array}\right) \left(\begin{array}{c}E_{  L}^+\\E_{  R}^+\end{array}\right)_{z=0}\,.
    \label{copropagatingsolution for fields}
\end{equation}
On noting that $2\kappa=k_0\left(\tilde{n}-n_{ {b}}\right)$ and applying the condition $k_0\alpha=-hp$, the on-resonance ($\delta_{  c}=0$) representation of Eq.\ \eqref{copropagatingsolution for fields} coincides with Eq.\ \eqref{eq Ofs FEMA}. In this instance, coupled wave theory, despite being an approximate method, yields the exact result.

\par Examining the on-resonance approximate expression, $R^{\rm Peak}=\tanh^2{\left(\kappa L\right)}$ and for a particular handedness, say right, for RCP, we have $k_0^{ {R}}={p}/{\left(\bar{n}-a\right)}$, while for LCP, $k_0^{ {L}}=-{p}/{\left(\bar{n}+a\right)}$. Then, 
\begin{equation}
    R_{  RR}^{\rm Peak}=\tanh^2{\left(\frac{p}{\bar{n}-{\a}}\frac{\mu\delta\varepsilon}{2\bar{n}}L\right)}\,,
\label{eq R resonance 2}
\end{equation}
where it is evident that increasing ${\alpha}$, increases $R_{  RR}^{ \rm Peak}$. On the contrary, for a left-handed medium, increasing ${\alpha}$ has the opposite impact on $R_{  LL}^{\rm Peak}$. Therefore, Eq.\ \eqref{eq R resonance 2} clearly demonstrates the monotonic of $R_{  RR}^{\rm Peak}$ with $\alpha$, as observed in \cite{Sherwin2003a, Sherwin2003b}.

\par Finally, as it is seen in \cite{Sherwin2003a},  optical activity also affects the resonance's bandwidth. This can be estimated as ${\mathrm{\Delta\lambda}}_0={\rm  Re}\left(\delta\bar{n}\right)L_{ {p}}$. Considering the optical rotation roughly proportional to the square of $\delta\bar{n}$ \cite{Hodgkinson2000}, we have $ {\mathrm{\Delta\lambda}}_0 \sim {\alpha}^2L_{ {p}}$ which qualitatively explains \cite[Eq. (40)]{Sherwin2003a}. 
\begin{acknowledgments}
The authors would like to acknowledge fruitful discussion with Dr. K. Weir of the Department of Physics at Imperial College London.
\end{acknowledgments}

\bibliography{sorsamp.bib}

\begin{thebibliography}{40}%
\makeatletter
\providecommand \@ifxundefined [1]{%
 \@ifx{#1\undefined}
}%
\providecommand \@ifnum [1]{%
 \ifnum #1\expandafter \@firstoftwo
 \else \expandafter \@secondoftwo
 \fi
}%
\providecommand \@ifx [1]{%
 \ifx #1\expandafter \@firstoftwo
 \else \expandafter \@secondoftwo
 \fi
}%
\providecommand \natexlab [1]{#1}%
\providecommand \enquote  [1]{``#1''}%
\providecommand \bibnamefont  [1]{#1}%
\providecommand \bibfnamefont [1]{#1}%
\providecommand \citenamefont [1]{#1}%
\providecommand \href@noop [0]{\@secondoftwo}%
\providecommand \href [0]{\begingroup \@sanitize@url \@href}%
\providecommand \@href[1]{\@@startlink{#1}\@@href}%
\providecommand \@@href[1]{\endgroup#1\@@endlink}%
\providecommand \@sanitize@url [0]{\catcode `\\12\catcode `\$12\catcode
  `\&12\catcode `\#12\catcode `\^12\catcode `\_12\catcode `\%12\relax}%
\providecommand \@@startlink[1]{}%
\providecommand \@@endlink[0]{}%
\providecommand \url  [0]{\begingroup\@sanitize@url \@url }%
\providecommand \@url [1]{\endgroup\@href {#1}{\urlprefix }}%
\providecommand \urlprefix  [0]{URL }%
\providecommand \Eprint [0]{\href }%
\providecommand \doibase [0]{https://doi.org/}%
\providecommand \selectlanguage [0]{\@gobble}%
\providecommand \bibinfo  [0]{\@secondoftwo}%
\providecommand \bibfield  [0]{\@secondoftwo}%
\providecommand \translation [1]{[#1]}%
\providecommand \BibitemOpen [0]{}%
\providecommand \bibitemStop [0]{}%
\providecommand \bibitemNoStop [0]{.\EOS\space}%
\providecommand \EOS [0]{\spacefactor3000\relax}%
\providecommand \BibitemShut  [1]{\csname bibitem#1\endcsname}%
\let\auto@bib@innerbib\@empty
\bibitem [{\citenamefont {Michelson}(1911)}]{Michelson1911}%
  \BibitemOpen
  \bibfield  {author} {\bibinfo {author} {\bibfnamefont {A.~A.}\ \bibnamefont
  {Michelson}},\ }\bibfield  {title} {\bibinfo {title} {On metallic colouring
  in birds and insects},\ }\href {https://doi.org/10.1080/14786440408637061}
  {\bibfield  {journal} {\bibinfo  {journal} {Lond. Edinb. Dublin Philos. Mag.
  J. Sci.}\ }\textbf {\bibinfo {volume} {21}},\ \bibinfo {pages} {554}
  (\bibinfo {year} {1911})}\BibitemShut {NoStop}%
\bibitem [{\citenamefont {Kang}\ \emph {et~al.}(2015)\citenamefont {Kang},
  \citenamefont {Lan}, \citenamefont {Cui}, \citenamefont {Rodrigues},
  \citenamefont {Liu}, \citenamefont {Werner},\ and\ \citenamefont
  {Cai}}]{Kang2015}%
  \BibitemOpen
  \bibfield  {author} {\bibinfo {author} {\bibfnamefont {L.}~\bibnamefont
  {Kang}}, \bibinfo {author} {\bibfnamefont {S.}~\bibnamefont {Lan}}, \bibinfo
  {author} {\bibfnamefont {Y.}~\bibnamefont {Cui}}, \bibinfo {author}
  {\bibfnamefont {S.~P.}\ \bibnamefont {Rodrigues}}, \bibinfo {author}
  {\bibfnamefont {Y.}~\bibnamefont {Liu}}, \bibinfo {author} {\bibfnamefont
  {D.~H.}\ \bibnamefont {Werner}},\ and\ \bibinfo {author} {\bibfnamefont
  {W.}~\bibnamefont {Cai}},\ }\bibfield  {title} {\bibinfo {title} {An active
  metamaterial platform for chiral responsive optoelectronics},\ }\href
  {https://doi.org/10.1002/adma.201501930} {\bibfield  {journal} {\bibinfo
  {journal} {Adv. Mater.}\ }\textbf {\bibinfo {volume} {27}},\ \bibinfo {pages}
  {4377} (\bibinfo {year} {2015})}\BibitemShut {NoStop}%
\bibitem [{\citenamefont {Yoo}\ and\ \citenamefont {Park}(2019)}]{Yoo2019}%
  \BibitemOpen
  \bibfield  {author} {\bibinfo {author} {\bibfnamefont {S.}~\bibnamefont
  {Yoo}}\ and\ \bibinfo {author} {\bibfnamefont {Q.~H.}\ \bibnamefont {Park}},\
  }\bibfield  {title} {\bibinfo {title} {Metamaterials and chiral sensing: a
  review of fundamentals and applications},\ }\href
  {https://doi.org/10.1515/nanoph-2018-0167} {\bibfield  {journal} {\bibinfo
  {journal} {Nanophotonics}\ }\textbf {\bibinfo {volume} {8}},\ \bibinfo
  {pages} {249} (\bibinfo {year} {2019})}\BibitemShut {NoStop}%
\bibitem [{\citenamefont {Lakhwani}\ and\ \citenamefont
  {Meskers}(2012)}]{Lakhwani2012}%
  \BibitemOpen
  \bibfield  {author} {\bibinfo {author} {\bibfnamefont {G.}~\bibnamefont
  {Lakhwani}}\ and\ \bibinfo {author} {\bibfnamefont {S.~C.~J.}\ \bibnamefont
  {Meskers}},\ }\bibfield  {title} {\bibinfo {title} {Insights from chiral
  polyfluorene on the unification of molecular exciton and cholesteric liquid
  crystal theories for chiroptical phenomena},\ }\href
  {https://doi.org/10.1021/jp209893h} {\bibfield  {journal} {\bibinfo
  {journal} {J. Phys. Chem. A}\ }\textbf {\bibinfo {volume} {116}},\ \bibinfo
  {pages} {1121} (\bibinfo {year} {2012})}\BibitemShut {NoStop}%
\bibitem [{\citenamefont {Robbie}\ \emph {et~al.}(1996)\citenamefont {Robbie},
  \citenamefont {Brett},\ and\ \citenamefont {Lakhtakia}}]{Robbie1996}%
  \BibitemOpen
  \bibfield  {author} {\bibinfo {author} {\bibfnamefont {K.}~\bibnamefont
  {Robbie}}, \bibinfo {author} {\bibfnamefont {M.~J.}\ \bibnamefont {Brett}},\
  and\ \bibinfo {author} {\bibfnamefont {A.}~\bibnamefont {Lakhtakia}},\
  }\bibfield  {title} {\bibinfo {title} {Chiral sculptured thin films},\ }\href
  {https://doi.org/10.1038/384616a0} {\bibfield  {journal} {\bibinfo  {journal}
  {Nature}\ }\textbf {\bibinfo {volume} {384}},\ \bibinfo {pages} {616}
  (\bibinfo {year} {1996})}\BibitemShut {NoStop}%
\bibitem [{\citenamefont {Arteaga}(2016)}]{Arteaga2016}%
  \BibitemOpen
  \bibfield  {author} {\bibinfo {author} {\bibfnamefont {O.}~\bibnamefont
  {Arteaga}},\ }\bibfield  {title} {\bibinfo {title} {Natural optical activity
  vs circular bragg reflection studied by mueller matrix ellipsometry},\ }\href
  {https://doi.org/10.1016/j.tsf.2016.01.012} {\bibfield  {journal} {\bibinfo
  {journal} {Thin Solid Films}\ }\textbf {\bibinfo {volume} {617}},\ \bibinfo
  {pages} {14} (\bibinfo {year} {2016})}\BibitemShut {NoStop}%
\bibitem [{\citenamefont {Wade}\ \emph {et~al.}(2020)\citenamefont {Wade},
  \citenamefont {Hilfiker}, \citenamefont {Brandt}, \citenamefont
  {Liiro~Peluso}, \citenamefont {Wan}, \citenamefont {Shi}, \citenamefont
  {Salerno}, \citenamefont {Ryan}, \citenamefont {Schoeche}, \citenamefont
  {Arteaga}, \citenamefont {Javorfi}, \citenamefont {Siligardi}, \citenamefont
  {Wang}, \citenamefont {Amabilino}, \citenamefont {Beton}, \citenamefont
  {Campbell},\ and\ \citenamefont {Fuchter}}]{Campbell2020}%
  \BibitemOpen
  \bibfield  {author} {\bibinfo {author} {\bibfnamefont {J.}~\bibnamefont
  {Wade}}, \bibinfo {author} {\bibfnamefont {J.}~\bibnamefont {Hilfiker}},
  \bibinfo {author} {\bibfnamefont {J.}~\bibnamefont {Brandt}}, \bibinfo
  {author} {\bibfnamefont {L.}~\bibnamefont {Liiro~Peluso}}, \bibinfo {author}
  {\bibfnamefont {L.}~\bibnamefont {Wan}}, \bibinfo {author} {\bibfnamefont
  {X.}~\bibnamefont {Shi}}, \bibinfo {author} {\bibfnamefont {F.}~\bibnamefont
  {Salerno}}, \bibinfo {author} {\bibfnamefont {S.}~\bibnamefont {Ryan}},
  \bibinfo {author} {\bibfnamefont {S.}~\bibnamefont {Schoeche}}, \bibinfo
  {author} {\bibfnamefont {O.}~\bibnamefont {Arteaga}}, \bibinfo {author}
  {\bibfnamefont {T.}~\bibnamefont {Javorfi}}, \bibinfo {author} {\bibfnamefont
  {G.}~\bibnamefont {Siligardi}}, \bibinfo {author} {\bibfnamefont
  {C.}~\bibnamefont {Wang}}, \bibinfo {author} {\bibfnamefont {D.}~\bibnamefont
  {Amabilino}}, \bibinfo {author} {\bibfnamefont {P.}~\bibnamefont {Beton}},
  \bibinfo {author} {\bibfnamefont {A.}~\bibnamefont {Campbell}},\ and\
  \bibinfo {author} {\bibfnamefont {M.}~\bibnamefont {Fuchter}},\ }\bibfield
  {title} {\bibinfo {title} {Natural optical activity as the origin of the
  large chiroptical properties in $\pi$-conjugated polymer thin films},\ }\href
  {https://doi.org/10.1038/s41467-020-19951-y} {\bibfield  {journal} {\bibinfo
  {journal} {Nature Communications}\ }\textbf {\bibinfo {volume} {11}},\
  \bibinfo {pages} {6137} (\bibinfo {year} {2020})}\BibitemShut {NoStop}%
\bibitem [{\citenamefont {Han}\ \emph {et~al.}(2018)\citenamefont {Han},
  \citenamefont {Guo}, \citenamefont {Lu}, \citenamefont {Liu}, \citenamefont
  {Zhao},\ and\ \citenamefont {Huang}}]{Han2018}%
  \BibitemOpen
  \bibfield  {author} {\bibinfo {author} {\bibfnamefont {J.~I.}\ \bibnamefont
  {Han}}, \bibinfo {author} {\bibfnamefont {S.}~\bibnamefont {Guo}}, \bibinfo
  {author} {\bibfnamefont {H.}~\bibnamefont {Lu}}, \bibinfo {author}
  {\bibfnamefont {S.}~\bibnamefont {Liu}}, \bibinfo {author} {\bibfnamefont
  {Q.}~\bibnamefont {Zhao}},\ and\ \bibinfo {author} {\bibfnamefont
  {W.}~\bibnamefont {Huang}},\ }\bibfield  {title} {\bibinfo {title} {Recent
  progress on circularly polarized luminescent materials for organic
  optoelectronic devices},\ }\href {https://doi.org/10.1002/adom.201800538}
  {\bibfield  {journal} {\bibinfo  {journal} {Adv. Opt. Mater.}\ }\textbf
  {\bibinfo {volume} {6}},\ \bibinfo {pages} {1800538} (\bibinfo {year}
  {2018})}\BibitemShut {NoStop}%
\bibitem [{\citenamefont {Li}\ \emph {et~al.}(2015)\citenamefont {Li},
  \citenamefont {Jing}, \citenamefont {Liu}, \citenamefont {Zhao},
  \citenamefont {Shi}, \citenamefont {Tang}, \citenamefont {Zheng},\ and\
  \citenamefont {Zuo}}]{Li2015}%
  \BibitemOpen
  \bibfield  {author} {\bibinfo {author} {\bibfnamefont {T.~Y.}\ \bibnamefont
  {Li}}, \bibinfo {author} {\bibfnamefont {Y.}~\bibnamefont {Jing}}, \bibinfo
  {author} {\bibfnamefont {X.}~\bibnamefont {Liu}}, \bibinfo {author}
  {\bibfnamefont {Y.}~\bibnamefont {Zhao}}, \bibinfo {author} {\bibfnamefont
  {L.}~\bibnamefont {Shi}}, \bibinfo {author} {\bibfnamefont {Z.}~\bibnamefont
  {Tang}}, \bibinfo {author} {\bibfnamefont {Y.~X.}\ \bibnamefont {Zheng}},\
  and\ \bibinfo {author} {\bibfnamefont {J.~L.}\ \bibnamefont {Zuo}},\
  }\bibfield  {title} {\bibinfo {title} {Circularly polarised phosphorescent
  photoluminescence and electroluminescence of iridium complexes},\ }\href
  {https://doi.org/10.1038/srep14912} {\bibfield  {journal} {\bibinfo
  {journal} {Sci. Rep.}\ }\textbf {\bibinfo {volume} {5}},\ \bibinfo {pages}
  {1} (\bibinfo {year} {2015})}\BibitemShut {NoStop}%
\bibitem [{\citenamefont {Yang}\ \emph {et~al.}(2013)\citenamefont {Yang},
  \citenamefont {Da~Costa}, \citenamefont {Fuchter},\ and\ \citenamefont
  {Campbell}}]{Yang2013}%
  \BibitemOpen
  \bibfield  {author} {\bibinfo {author} {\bibfnamefont {Y.}~\bibnamefont
  {Yang}}, \bibinfo {author} {\bibfnamefont {R.~C.}\ \bibnamefont {Da~Costa}},
  \bibinfo {author} {\bibfnamefont {M.~J.}\ \bibnamefont {Fuchter}},\ and\
  \bibinfo {author} {\bibfnamefont {A.~J.}\ \bibnamefont {Campbell}},\
  }\bibfield  {title} {\bibinfo {title} {Circularly polarized light detection
  by a chiral organic semiconductor transistor},\ }\href
  {https://doi.org/10.1038/nphoton.2013.176} {\bibfield  {journal} {\bibinfo
  {journal} {Nat. Photonics}\ }\textbf {\bibinfo {volume} {7}},\ \bibinfo
  {pages} {634} (\bibinfo {year} {2013})}\BibitemShut {NoStop}%
\bibitem [{\citenamefont {McCall}\ and\ \citenamefont
  {Lakhtakia}(2000)}]{McCall2000}%
  \BibitemOpen
  \bibfield  {author} {\bibinfo {author} {\bibfnamefont {M.~W.}\ \bibnamefont
  {McCall}}\ and\ \bibinfo {author} {\bibfnamefont {A.}~\bibnamefont
  {Lakhtakia}},\ }\bibfield  {title} {\bibinfo {title} {Development and
  assessment of coupled wave theory of axial propagation in thin-film
  helicoidal bianisotropic media. part 1: Reflectances and transmittances},\
  }\href {https://doi.org/10.1080/09500340008233400} {\bibfield  {journal}
  {\bibinfo  {journal} {J. Mod. Opt.}\ }\textbf {\bibinfo {volume} {47}},\
  \bibinfo {pages} {973} (\bibinfo {year} {2000})}\BibitemShut {NoStop}%
\bibitem [{\citenamefont {McCall}\ and\ \citenamefont
  {Lakhtakia}(2001)}]{McCall2001}%
  \BibitemOpen
  \bibfield  {author} {\bibinfo {author} {\bibfnamefont {M.~W.}\ \bibnamefont
  {McCall}}\ and\ \bibinfo {author} {\bibfnamefont {A.}~\bibnamefont
  {Lakhtakia}},\ }\bibfield  {title} {\bibinfo {title} {Development and
  assessment of coupled wave theory of axial propagation in thin-film
  helicoidal bi-anisotropic media. part 2: Dichroisms, ellipticity
  transformation and optical rotation},\ }\href
  {https://www.tandfonline.com/doi/abs/10.1080/09500340108235161} {\bibfield
  {journal} {\bibinfo  {journal} {J. Mod. Opt.}\ }\textbf {\bibinfo {volume}
  {48}},\ \bibinfo {pages} {143} (\bibinfo {year} {2001})}\BibitemShut
  {NoStop}%
\bibitem [{\citenamefont {McCall}\ and\ \citenamefont
  {Lakhtakia}(2004)}]{McCall2004}%
  \BibitemOpen
  \bibfield  {author} {\bibinfo {author} {\bibfnamefont {M.~W.}\ \bibnamefont
  {McCall}}\ and\ \bibinfo {author} {\bibfnamefont {A.}~\bibnamefont
  {Lakhtakia}},\ }\bibfield  {title} {\bibinfo {title} {Explicit expressions
  for spectral remittances of axially excited chiral sculptured thin films},\
  }\href {https://doi.org/10.1080/09500340408234596} {\bibfield  {journal}
  {\bibinfo  {journal} {J. Mod. Opt.}\ }\textbf {\bibinfo {volume} {51}},\
  \bibinfo {pages} {111} (\bibinfo {year} {2004})}\BibitemShut {NoStop}%
\bibitem [{\citenamefont {Lakhtakia}\ and\ \citenamefont
  {Messier}(2005)}]{LakhtakiaBook2005}%
  \BibitemOpen
  \bibfield  {author} {\bibinfo {author} {\bibfnamefont {A.}~\bibnamefont
  {Lakhtakia}}\ and\ \bibinfo {author} {\bibfnamefont {R.}~\bibnamefont
  {Messier}},\ }\href {https://doi.org/10.1016/j.ijleo.2006.03.008} {\emph
  {\bibinfo {title} {Sculptured Thin Films: Nanoengineered Morphology and
  Optics}}}\ (\bibinfo  {publisher} {SPIE Optical Engineering Press},\ \bibinfo
  {year} {2005})\BibitemShut {NoStop}%
\bibitem [{\citenamefont {McCall}(2009{\natexlab{a}})}]{McCall2009}%
  \BibitemOpen
  \bibfield  {author} {\bibinfo {author} {\bibfnamefont {M.~W.}\ \bibnamefont
  {McCall}},\ }\bibfield  {title} {\bibinfo {title} {Simplified theory of axial
  propagation through structurally chiral media},\ }\href
  {https://doi.org/10.1088/1464-4258/11/7/074006} {\bibfield  {journal}
  {\bibinfo  {journal} {J. Opt.}\ }\textbf {\bibinfo {volume} {11}},\ \bibinfo
  {pages} {074006} (\bibinfo {year} {2009}{\natexlab{a}})}\BibitemShut
  {NoStop}%
\bibitem [{\citenamefont {Mackay}\ and\ \citenamefont
  {Lakhtakia}(2010{\natexlab{a}})}]{Mackay2010sensing}%
  \BibitemOpen
  \bibfield  {author} {\bibinfo {author} {\bibfnamefont {T.~G.}\ \bibnamefont
  {Mackay}}\ and\ \bibinfo {author} {\bibfnamefont {A.}~\bibnamefont
  {Lakhtakia}},\ }\bibfield  {title} {\bibinfo {title} {On the theory of
  optical sensing via infiltration of sculptured thin films},\ }in\ \href
  {https://doi.org/10.1117/12.859655} {\emph {\bibinfo {booktitle}
  {Nanostructured Thin Films III}}},\ Vol.\ \bibinfo {volume} {7766}\ (\bibinfo
  {organization} {SPIE},\ \bibinfo {year} {2010})\ pp.\ \bibinfo {pages}
  {117--26}\BibitemShut {NoStop}%
\bibitem [{\citenamefont {Lakhtakia}(2000)}]{Lakhtakia2000}%
  \BibitemOpen
  \bibfield  {author} {\bibinfo {author} {\bibfnamefont {A.}~\bibnamefont
  {Lakhtakia}},\ }\bibfield  {title} {\bibinfo {title} {On percolation and
  circular bragg phenomenon in metallic, helicoidally periodic, sculptured thin
  films},\ }\href
  {https://onlinelibrary.wiley.com/doi/pdf/10.1002/%28SICI%291098-2760%2820000220%2924%3A4%3C239%3A%3AAID-MOP10%3E3.0.CO%3B2-%23}
  {\bibfield  {journal} {\bibinfo  {journal} {Microw. Opt. Technol. Lett.}\
  }\textbf {\bibinfo {volume} {24}},\ \bibinfo {pages} {239} (\bibinfo {year}
  {2000})}\BibitemShut {NoStop}%
\bibitem [{\citenamefont {Sorge}\ \emph {et~al.}(2006)\citenamefont {Sorge},
  \citenamefont {van Popta}, \citenamefont {Sit},\ and\ \citenamefont
  {Brett}}]{Sorge2006}%
  \BibitemOpen
  \bibfield  {author} {\bibinfo {author} {\bibfnamefont {J.~B.}\ \bibnamefont
  {Sorge}}, \bibinfo {author} {\bibfnamefont {A.~C.}\ \bibnamefont {van
  Popta}}, \bibinfo {author} {\bibfnamefont {J.~C.}\ \bibnamefont {Sit}},\ and\
  \bibinfo {author} {\bibfnamefont {M.~J.}\ \bibnamefont {Brett}},\ }\bibfield
  {title} {\bibinfo {title} {Circular birefringence dependence on chiral film
  porosity},\ }\href {https://doi.org/10.1364/OE.14.010550} {\bibfield
  {journal} {\bibinfo  {journal} {Opt. Express}\ }\textbf {\bibinfo {volume}
  {14}},\ \bibinfo {pages} {10550} (\bibinfo {year} {2006})}\BibitemShut
  {NoStop}%
\bibitem [{\citenamefont {Shirin}\ \emph {et~al.}(2020)\citenamefont {Shirin},
  \citenamefont {Madani},\ and\ \citenamefont {Entezar}}]{Shirin2020}%
  \BibitemOpen
  \bibfield  {author} {\bibinfo {author} {\bibfnamefont {S.}~\bibnamefont
  {Shirin}}, \bibinfo {author} {\bibfnamefont {A.}~\bibnamefont {Madani}},\
  and\ \bibinfo {author} {\bibfnamefont {S.~R.}\ \bibnamefont {Entezar}},\
  }\bibfield  {title} {\bibinfo {title} {Tunable lateral shift of the reflected
  optical beams from a nanocomposite structurally chiral medium},\ }\href
  {https://doi.org/10.1016/j.optmat.2020.110026} {\bibfield  {journal}
  {\bibinfo  {journal} {Opt. Mater.}\ }\textbf {\bibinfo {volume} {107}},\
  \bibinfo {pages} {110026} (\bibinfo {year} {2020})}\BibitemShut {NoStop}%
\bibitem [{\citenamefont {Pursel}\ \emph {et~al.}(2006)\citenamefont {Pursel},
  \citenamefont {Horn},\ and\ \citenamefont {Lakhtakia}}]{Pursel2006}%
  \BibitemOpen
  \bibfield  {author} {\bibinfo {author} {\bibfnamefont {S.~M.}\ \bibnamefont
  {Pursel}}, \bibinfo {author} {\bibfnamefont {M.~W.}\ \bibnamefont {Horn}},\
  and\ \bibinfo {author} {\bibfnamefont {A.}~\bibnamefont {Lakhtakia}},\
  }\bibfield  {title} {\bibinfo {title} {Blue-shifting of circular bragg
  phenomenon by annealing of chiral sculptured thin films},\ }\href
  {https://doi.org/10.1364/OE.14.008001} {\bibfield  {journal} {\bibinfo
  {journal} {Opt. Express}\ }\textbf {\bibinfo {volume} {14}},\ \bibinfo
  {pages} {8001} (\bibinfo {year} {2006})}\BibitemShut {NoStop}%
\bibitem [{\citenamefont {Mackay}\ and\ \citenamefont
  {Lakhtakia}(2010{\natexlab{b}})}]{Mackay2010}%
  \BibitemOpen
  \bibfield  {author} {\bibinfo {author} {\bibfnamefont {T.~G.}\ \bibnamefont
  {Mackay}}\ and\ \bibinfo {author} {\bibfnamefont {A.}~\bibnamefont
  {Lakhtakia}},\ }\bibfield  {title} {\bibinfo {title} {Empirical model of
  optical sensing via spectral shift of circular bragg phenomenon},\ }\href
  {https://doi.org/10.1109/JPHOT.2010.2042589} {\bibfield  {journal} {\bibinfo
  {journal} {IEEE Photonics J.}\ }\textbf {\bibinfo {volume} {2}},\ \bibinfo
  {pages} {92} (\bibinfo {year} {2010}{\natexlab{b}})}\BibitemShut {NoStop}%
\bibitem [{\citenamefont {Lakhtakia}(2001)}]{Lakhtakia2001}%
  \BibitemOpen
  \bibfield  {author} {\bibinfo {author} {\bibfnamefont {A.}~\bibnamefont
  {Lakhtakia}},\ }\bibfield  {title} {\bibinfo {title} {Enhancement of optical
  activity of chiral sculptured thin films by suitable infiltration of void
  regions},\ }\href {https://doi.org/10.1078/0030-4026-00024} {\bibfield
  {journal} {\bibinfo  {journal} {Optik (Stuttg.)}\ }\textbf {\bibinfo {volume}
  {112}},\ \bibinfo {pages} {145} (\bibinfo {year} {2001})}\BibitemShut
  {NoStop}%
\bibitem [{\citenamefont {Hodgkinson}\ \emph {et~al.}(2000)\citenamefont
  {Hodgkinson}, \citenamefont {Wu}, \citenamefont {Knight}, \citenamefont
  {Lakhtakia},\ and\ \citenamefont {Robbie}}]{Hodgkinson2000}%
  \BibitemOpen
  \bibfield  {author} {\bibinfo {author} {\bibfnamefont {I.}~\bibnamefont
  {Hodgkinson}}, \bibinfo {author} {\bibfnamefont {Q.~H.}\ \bibnamefont {Wu}},
  \bibinfo {author} {\bibfnamefont {B.}~\bibnamefont {Knight}}, \bibinfo
  {author} {\bibfnamefont {A.}~\bibnamefont {Lakhtakia}},\ and\ \bibinfo
  {author} {\bibfnamefont {K.}~\bibnamefont {Robbie}},\ }\bibfield  {title}
  {\bibinfo {title} {Vacuum deposition of chiral sculptured thin films with
  high optical activity},\ }\href {https://doi.org/10.1364/AO.39.000642}
  {\bibfield  {journal} {\bibinfo  {journal} {Appl. Opt.}\ }\textbf {\bibinfo
  {volume} {39}},\ \bibinfo {pages} {642} (\bibinfo {year} {2000})}\BibitemShut
  {NoStop}%
\bibitem [{\citenamefont {Sherwin}\ and\ \citenamefont
  {Lakhtakia}(2002)}]{Sherwin2003a}%
  \BibitemOpen
  \bibfield  {author} {\bibinfo {author} {\bibfnamefont {J.~A.}\ \bibnamefont
  {Sherwin}}\ and\ \bibinfo {author} {\bibfnamefont {A.}~\bibnamefont
  {Lakhtakia}},\ }\bibfield  {title} {\bibinfo {title} {Nominal model for the
  optical response of a chiral sculptured thin film infiltrated with an
  isotropic chiral fluid},\ }\href
  {https://doi.org/10.1016/S0030-4018(02)02180-6} {\bibfield  {journal}
  {\bibinfo  {journal} {Opt. Commun.}\ }\textbf {\bibinfo {volume} {214}},\
  \bibinfo {pages} {231} (\bibinfo {year} {2002})}\BibitemShut {NoStop}%
\bibitem [{\citenamefont {Sherwin}\ and\ \citenamefont
  {Lakhtakia}(2003)}]{Sherwin2003b}%
  \BibitemOpen
  \bibfield  {author} {\bibinfo {author} {\bibfnamefont {J.~A.}\ \bibnamefont
  {Sherwin}}\ and\ \bibinfo {author} {\bibfnamefont {A.}~\bibnamefont
  {Lakhtakia}},\ }\bibfield  {title} {\bibinfo {title} {Nominal model for the
  optical response of a chiral sculptured thin film infiltrated by an isotropic
  chiral fluid-oblique incidence},\ }\href
  {https://doi.org/10.1016/S0030-4018(03)01609-2} {\bibfield  {journal}
  {\bibinfo  {journal} {Opt. Commun.}\ }\textbf {\bibinfo {volume} {222}},\
  \bibinfo {pages} {305} (\bibinfo {year} {2003})}\BibitemShut {NoStop}%
\bibitem [{\citenamefont {Hodgkinson}\ \emph {et~al.}(2004)\citenamefont
  {Hodgkinson}, \citenamefont {Lakhtakia}, \citenamefont {Wu}, \citenamefont
  {De~Silva},\ and\ \citenamefont {McCall}}]{Hodgkinson2004ambichiral}%
  \BibitemOpen
  \bibfield  {author} {\bibinfo {author} {\bibfnamefont {I.~J.}\ \bibnamefont
  {Hodgkinson}}, \bibinfo {author} {\bibfnamefont {A.}~\bibnamefont
  {Lakhtakia}}, \bibinfo {author} {\bibfnamefont {Q.~h.}\ \bibnamefont {Wu}},
  \bibinfo {author} {\bibfnamefont {L.}~\bibnamefont {De~Silva}},\ and\
  \bibinfo {author} {\bibfnamefont {M.~W.}\ \bibnamefont {McCall}},\ }\bibfield
   {title} {\bibinfo {title} {Ambichiral, equichiral and finely chiral layered
  structures},\ }\href {https://doi.org/10.1016/j.optcom.2004.06.005}
  {\bibfield  {journal} {\bibinfo  {journal} {Opt. Commun.}\ }\textbf {\bibinfo
  {volume} {239}},\ \bibinfo {pages} {353} (\bibinfo {year}
  {2004})}\BibitemShut {NoStop}%
\bibitem [{\citenamefont {van Popta}\ \emph {et~al.}(2005)\citenamefont {van
  Popta}, \citenamefont {Brett},\ and\ \citenamefont {Sit}}]{vanPopta2005}%
  \BibitemOpen
  \bibfield  {author} {\bibinfo {author} {\bibfnamefont {A.~C.}\ \bibnamefont
  {van Popta}}, \bibinfo {author} {\bibfnamefont {M.~J.}\ \bibnamefont
  {Brett}},\ and\ \bibinfo {author} {\bibfnamefont {J.~C.}\ \bibnamefont
  {Sit}},\ }\bibfield  {title} {\bibinfo {title} {Double-handed circular bragg
  phenomena in polygonal helix thin films},\ }\href
  {https://doi.org/10.1063/1.2115092} {\bibfield  {journal} {\bibinfo
  {journal} {J. Appl. Phys.}\ }\textbf {\bibinfo {volume} {98}},\ \bibinfo
  {pages} {083517} (\bibinfo {year} {2005})}\BibitemShut {NoStop}%
\bibitem [{\citenamefont {Pendry}(2004)}]{Pendry2004}%
  \BibitemOpen
  \bibfield  {author} {\bibinfo {author} {\bibfnamefont {J.~B.}\ \bibnamefont
  {Pendry}},\ }\bibfield  {title} {\bibinfo {title} {A chiral route to negative
  refraction},\ }\href {https://doi.org/10.1126/science.1104467} {\bibfield
  {journal} {\bibinfo  {journal} {Science}\ }\textbf {\bibinfo {volume}
  {306}},\ \bibinfo {pages} {1353} (\bibinfo {year} {2004})}\BibitemShut
  {NoStop}%
\bibitem [{\citenamefont {Monzon}\ and\ \citenamefont
  {Forester}(2005)}]{Monzon2005}%
  \BibitemOpen
  \bibfield  {author} {\bibinfo {author} {\bibfnamefont {C.}~\bibnamefont
  {Monzon}}\ and\ \bibinfo {author} {\bibfnamefont {D.~W.}\ \bibnamefont
  {Forester}},\ }\bibfield  {title} {\bibinfo {title} {Negative refraction and
  focusing of circularly polarized waves in optically active media},\ }\href
  {https://doi.org/10.1103/PhysRevLett.95.123904} {\bibfield  {journal}
  {\bibinfo  {journal} {Phys.\ Rev.\ Lett.}\ }\textbf {\bibinfo {volume}
  {95}},\ \bibinfo {pages} {123904} (\bibinfo {year} {2005})}\BibitemShut
  {NoStop}%
\bibitem [{\citenamefont {Zhang}\ \emph {et~al.}(2009)\citenamefont {Zhang},
  \citenamefont {Park}, \citenamefont {Li}, \citenamefont {Lu}, \citenamefont
  {Zhang},\ and\ \citenamefont {Zhang}}]{Zhang2009}%
  \BibitemOpen
  \bibfield  {author} {\bibinfo {author} {\bibfnamefont {S.}~\bibnamefont
  {Zhang}}, \bibinfo {author} {\bibfnamefont {Y.-S.}\ \bibnamefont {Park}},
  \bibinfo {author} {\bibfnamefont {J.}~\bibnamefont {Li}}, \bibinfo {author}
  {\bibfnamefont {X.}~\bibnamefont {Lu}}, \bibinfo {author} {\bibfnamefont
  {W.}~\bibnamefont {Zhang}},\ and\ \bibinfo {author} {\bibfnamefont
  {X.}~\bibnamefont {Zhang}},\ }\bibfield  {title} {\bibinfo {title} {Negative
  refractive index in chiral metamaterials},\ }\href
  {https://doi.org/10.1103/PhysRevLett.102.023901} {\bibfield  {journal}
  {\bibinfo  {journal} {Phys.\ Rev.\ Lett.}\ }\textbf {\bibinfo {volume}
  {102}},\ \bibinfo {pages} {023901} (\bibinfo {year} {2009})}\BibitemShut
  {NoStop}%
\bibitem [{\citenamefont {Zhou}\ \emph {et~al.}(2009)\citenamefont {Zhou},
  \citenamefont {Dong}, \citenamefont {Wang}, \citenamefont {Koschny},
  \citenamefont {Kafesaki},\ and\ \citenamefont {Soukoulis}}]{Kafesak2009}%
  \BibitemOpen
  \bibfield  {author} {\bibinfo {author} {\bibfnamefont {J.}~\bibnamefont
  {Zhou}}, \bibinfo {author} {\bibfnamefont {J.}~\bibnamefont {Dong}}, \bibinfo
  {author} {\bibfnamefont {B.}~\bibnamefont {Wang}}, \bibinfo {author}
  {\bibfnamefont {T.}~\bibnamefont {Koschny}}, \bibinfo {author} {\bibfnamefont
  {M.}~\bibnamefont {Kafesaki}},\ and\ \bibinfo {author} {\bibfnamefont
  {C.~M.}\ \bibnamefont {Soukoulis}},\ }\bibfield  {title} {\bibinfo {title}
  {Negative refractive index due to chirality},\ }\href
  {https://doi.org/10.1103/PhysRevB.79.121104} {\bibfield  {journal} {\bibinfo
  {journal} {Phys. Rev. B}\ }\textbf {\bibinfo {volume} {79}},\ \bibinfo
  {pages} {121104} (\bibinfo {year} {2009})}\BibitemShut {NoStop}%
\bibitem [{\citenamefont {Plum}\ \emph {et~al.}(2009)\citenamefont {Plum},
  \citenamefont {Zhou}, \citenamefont {Dong}, \citenamefont {Fedotov},
  \citenamefont {Koschny}, \citenamefont {Soukoulis},\ and\ \citenamefont
  {Zheludev}}]{Zheludev2009}%
  \BibitemOpen
  \bibfield  {author} {\bibinfo {author} {\bibfnamefont {E.}~\bibnamefont
  {Plum}}, \bibinfo {author} {\bibfnamefont {J.}~\bibnamefont {Zhou}}, \bibinfo
  {author} {\bibfnamefont {J.}~\bibnamefont {Dong}}, \bibinfo {author}
  {\bibfnamefont {V.~A.}\ \bibnamefont {Fedotov}}, \bibinfo {author}
  {\bibfnamefont {T.}~\bibnamefont {Koschny}}, \bibinfo {author} {\bibfnamefont
  {C.~M.}\ \bibnamefont {Soukoulis}},\ and\ \bibinfo {author} {\bibfnamefont
  {N.~I.}\ \bibnamefont {Zheludev}},\ }\bibfield  {title} {\bibinfo {title}
  {Metamaterial with negative index due to chirality},\ }\href
  {https://doi.org/10.1103/PhysRevB.79.035407} {\bibfield  {journal} {\bibinfo
  {journal} {Phys. Rev. B}\ }\textbf {\bibinfo {volume} {79}},\ \bibinfo
  {pages} {035407} (\bibinfo {year} {2009})}\BibitemShut {NoStop}%
\bibitem [{\citenamefont {Zarifi}\ \emph {et~al.}(2012)\citenamefont {Zarifi},
  \citenamefont {Soleimani},\ and\ \citenamefont {Nayyeri}}]{Zarifi2012}%
  \BibitemOpen
  \bibfield  {author} {\bibinfo {author} {\bibfnamefont {D.}~\bibnamefont
  {Zarifi}}, \bibinfo {author} {\bibfnamefont {M.}~\bibnamefont {Soleimani}},\
  and\ \bibinfo {author} {\bibfnamefont {V.}~\bibnamefont {Nayyeri}},\
  }\bibfield  {title} {\bibinfo {title} {Dual-and multiband chiral metamaterial
  structures with strong optical activity and negative refraction index},\
  }\href {https://doi.org/10.1109/LAWP.2012.2191261} {\bibfield  {journal}
  {\bibinfo  {journal} {IEEE Antennas Wirel. Propag. Lett.}\ }\textbf {\bibinfo
  {volume} {11}},\ \bibinfo {pages} {334} (\bibinfo {year} {2012})}\BibitemShut
  {NoStop}%
\bibitem [{\citenamefont {Lakhtakia}(2003)}]{Lakhtakia2003}%
  \BibitemOpen
  \bibfield  {author} {\bibinfo {author} {\bibfnamefont {A.}~\bibnamefont
  {Lakhtakia}},\ }\bibfield  {title} {\bibinfo {title} {Handedness reversal of
  circular bragg phenomenon due to negative real permittivity and
  permeability},\ }\href {https://doi.org/10.1364/OE.11.000716} {\bibfield
  {journal} {\bibinfo  {journal} {Opt. Express}\ }\textbf {\bibinfo {volume}
  {11}},\ \bibinfo {pages} {716} (\bibinfo {year} {2003})}\BibitemShut
  {NoStop}%
\bibitem [{\citenamefont {Bohren}(2003)}]{BohrenBook2003}%
  \BibitemOpen
  \bibfield  {author} {\bibinfo {author} {\bibfnamefont {C.~F.}\ \bibnamefont
  {Bohren}},\ }\href {https://spie.org/Publications/Book/504610} {\emph
  {\bibinfo {title} {Introduction to complex mediums for optics and
  electromagnetics}}},\ edited by\ \bibinfo {editor} {\bibfnamefont {W.~S.}\
  \bibnamefont {Weiglhofer}}\ and\ \bibinfo {editor} {\bibfnamefont
  {A.}~\bibnamefont {Lakhtakia}},\ Vol.\ \bibinfo {volume} {123}\ (\bibinfo
  {publisher} {SPIE Optical Engineering Press},\ \bibinfo {year} {2003})\
  Chap.\ \bibinfo {chapter} {3. Isotropic chiral materials}\BibitemShut
  {NoStop}%
\bibitem [{\citenamefont {Oseen}(1933)}]{Oseen1993}%
  \BibitemOpen
  \bibfield  {author} {\bibinfo {author} {\bibfnamefont {C.~W.}\ \bibnamefont
  {Oseen}},\ }\bibfield  {title} {\bibinfo {title} {The theory of liquid
  crystals},\ }\href {https://doi.org/10.1039/TF9332900883} {\bibfield
  {journal} {\bibinfo  {journal} {Trans. Faraday Soc.}\ }\textbf {\bibinfo
  {volume} {29}},\ \bibinfo {pages} {883} (\bibinfo {year} {1933})}\BibitemShut
  {NoStop}%
\bibitem [{\citenamefont {McCall}\ \emph {et~al.}(2015)\citenamefont {McCall},
  \citenamefont {Hodgkinson},\ and\ \citenamefont {Wu}}]{McCallBook2015}%
  \BibitemOpen
  \bibfield  {author} {\bibinfo {author} {\bibfnamefont {M.~W.}\ \bibnamefont
  {McCall}}, \bibinfo {author} {\bibfnamefont {I.}~\bibnamefont {Hodgkinson}},\
  and\ \bibinfo {author} {\bibfnamefont {Q.}~\bibnamefont {Wu}},\ }\href
  {https://doi.org/10.1142/p962} {\emph {\bibinfo {title} {Birefringent thin
  films and polarizing elements}}}\ (\bibinfo  {publisher} {London: Imperial
  College Press},\ \bibinfo {year} {2015})\ Chap.\ \bibinfo {chapter} {10:
  Continuum Methods}\BibitemShut {NoStop}%
\bibitem [{\citenamefont {Mohammadi-Baghaee}\ and\ \citenamefont
  {Rashed-Mohassel}(2016)}]{Mohammadi2016}%
  \BibitemOpen
  \bibfield  {author} {\bibinfo {author} {\bibfnamefont {R.}~\bibnamefont
  {Mohammadi-Baghaee}}\ and\ \bibinfo {author} {\bibfnamefont {J.}~\bibnamefont
  {Rashed-Mohassel}},\ }\bibfield  {title} {\bibinfo {title} {The chirality
  parameter for chiral chemical solutions},\ }\href
  {https://doi.org/10.1007/s10953-016-0496-4} {\bibfield  {journal} {\bibinfo
  {journal} {J. Solution. Chem.}\ }\textbf {\bibinfo {volume} {45}},\ \bibinfo
  {pages} {1171} (\bibinfo {year} {2016})}\BibitemShut {NoStop}%
\bibitem [{\citenamefont {Mackay}(2005)}]{Mackay2005}%
  \BibitemOpen
  \bibfield  {author} {\bibinfo {author} {\bibfnamefont {T.~G.}\ \bibnamefont
  {Mackay}},\ }\bibfield  {title} {\bibinfo {title} {Plane waves with negative
  phase velocity in isotropic chiral mediums},\ }\href
  {https://doi.org/10.1002/mop.20742} {\bibfield  {journal} {\bibinfo
  {journal} {Microw. Opt. Technol. Lett.}\ }\textbf {\bibinfo {volume} {45}},\
  \bibinfo {pages} {120} (\bibinfo {year} {2005})}\BibitemShut {NoStop}%
\bibitem [{\citenamefont
  {McCall}(2009{\natexlab{b}})}]{McCall2009negativerefraction}%
  \BibitemOpen
  \bibfield  {author} {\bibinfo {author} {\bibfnamefont {M.~W.}\ \bibnamefont
  {McCall}},\ }\bibfield  {title} {\bibinfo {title} {What is negative
  refraction?},\ }\href {https://doi.org/10.1080/09500340903324818} {\bibfield
  {journal} {\bibinfo  {journal} {J. Mod. Opt.}\ }\textbf {\bibinfo {volume}
  {56}},\ \bibinfo {pages} {1727} (\bibinfo {year}
  {2009}{\natexlab{b}})}\BibitemShut {NoStop}%
\end{thebibliography}%

\end{document}